%% file: Paper.tex
\documentclass[useAMS,a4,usenatbib]{mn2e}

\usepackage{amsmath}
\input{psfig.tex}
\input{macros.tex}
\begin{document}
%

\title[Clustering and Galaxy-Galaxy Lensing]
{Galaxy Clustering \& Galaxy-Galaxy Lensing:\\
A Promising Union to Constrain Cosmological Parameters}

\author[Cacciato et al.]
       {\parbox[t]{\textwidth}{
        Marcello Cacciato$^{1}$\thanks{International Max-Planck Research 
                       School Fellow \newline E-mail: cacciato@mpia.de}, 
        Frank C. van den Bosch$^{1}$, 
        Surhud More$^{1}$, 
        Ran Li$^{2,3}$, \\
        H.J. Mo$^{2}$, 
        Xiaohu Yang$^{4}$}\\
           \vspace*{3pt} \\
        $^1$Max-Planck-Institute for Astronomy, K\"onigstuhl 17, D-69117
            Heidelberg, Germany\\
        $^2$Department of Astronomy, University of Massachussetts, Amherst, 
            MA 01003-9305, USA\\
        $^3$Department of Astronomy, Peking University, Beijing 100871, China\\
        $^4$Shanghai Astronomical Observatory, Nandan Road 80, 
            Shanghai 200030, China}


\date{}

\pagerange{\pageref{firstpage}--\pageref{lastpage}}
\pubyear{2008}

\maketitle

\label{firstpage}


\begin{abstract}
  Galaxy  clustering and  galaxy-galaxy lensing  probe  the connection
  between galaxies and their dark matter haloes in complementary ways.
  Since the clustering of dark matter haloes depends on cosmology, the
  halo  occupation statistics  inferred from  the  observed clustering
  properties of  galaxies are degenerate  with the adopted  cosmology. 
  Consequently,  different cosmologies  imply  different mass-to-light
  ratios for dark matter  haloes.  Galaxy-galaxy lensing, which yields
  direct constraints on the actual mass-to-light ratios, can therefore
  be used to break this degeneracy, and thus to constrain cosmological
  parameters.   In this  paper we  establish the  link  between galaxy
  luminosity  and   dark  matter  halo  mass   using  the  conditional
  luminosity  function  (CLF), $\Phi(L|M){\rm  d}L$,  which gives  the
  number of galaxies with luminosities in the range $L \pm {\rm d}L/2$
  that reside in a halo of  mass $M$.  We constrain the CLF parameters
  using the  galaxy luminosity function and  the luminosity dependence
  of the  correlation lengths of  galaxies.  The resulting  CLF models
  are  used  to  predict  the  galaxy-galaxy lensing  signal.   For  a
  cosmology  that agrees  with constraints  from the  cosmic microwave
  background, i.e.   $(\Omega_{\rm m},\sigma_8) =  (0.238,0.734)$, the
  model accurately  fits the galaxy-galaxy lensing  data obtained from
  the   SDSS.    For  a   comparison   cosmology  with   $(\Omega_{\rm
    m},\sigma_8)  =  (0.3,0.9)$,  however,  we can  accurately  fit  the
  luminosity  function   and  clustering  properties   of  the  galaxy
  population, but the model predicts mass-to-light ratios that are too
  high,  resulting in  a  strong overprediction  of the  galaxy-galaxy
  lensing signal.  We conclude  that the combination of galaxy clustering and
  galaxy-galaxy lensing is  a powerful probe  of the galaxy-dark  matter connection,
  with  the  potential  to  yield tight  constraints  on  cosmological
  parameters.  Since  this method mainly probes  the mass distribution
  on  relatively small  (non-linear)  scales, it  is complementary  to
  constraints  obtained from the  galaxy power-spectrum,  which mainly
  probes the large-scale (linear) matter distribution.
\end{abstract}




\section{Introduction}
\label{sec:intro}

With   the  advent of large  galaxy  redshift  surveys, it  has become
possible to obtain accurate measurements of the clustering of galaxies
as a function of their properties, such  as luminosity, morphology and
color (e.g.  Guzzo \etal 2000; Norberg \etal  2001, 2002; Zehavi \etal
2005; Wang \etal 2007). Since galaxies are believed to form and reside
in  dark matter haloes, the clustering  strength of a given population
of galaxies can be compared to that of dark matter haloes as predicted
by  numerical simulations  or the  extended Press-Schechter formalism.
Such a comparison reveals a wealth of  information about the so-called
\textit{galaxy-dark matter connection} (e.g.    Jing, Mo  \&  B\"orner
1998; Peacock \& Smith 2000; Berlind \& Weinberg 2002; Yang, Mo \& van
den Bosch 2003; van  den Bosch, Yang \&  Mo 2003; van den Bosch  \etal
2007).

Unfortunately, this method of  constraining the link  between galaxies
and  dark matter  haloes  using  galaxy clustering   has one important
shortcoming: the halo occupation statistics inferred from the observed
clustering properties  depend on the cosmological  parameters adopted.
More precisely,  models based  on  different cosmologies can   fit the
clustering data    equally well by simply   relying  on different halo
occupation  statistics    or,  equivalently, different   mass-to-light
ratios. In  order to break this  degeneracy between cosmology and halo
occupation statistics   independent constraints  on  the mass-to-light
ratios are required  (e.g.   van den Bosch,   Mo \& Yang  2003; Tinker
\etal  2005).   One  method that  can  provide  these  constraints  is
galaxy-galaxy lensing (hereafter g-g lensing),  which probes the  mass
distributions (and  hence  the halo  masses)   around galaxies.   This
implies that  the combination of  clustering and  lensing in principle
holds the   potential to put   constraints  on cosmological parameters
(Seljak \etal 2005; Yoo \etal 2006).

The  first attempt  to  detect g-g  lensing  was  made by  Tyson \etal
(1984), but because of the relatively  poor quality of their data they
were unable to detect a  statistically  significant signal.  With  the
advent  of wider and deeper  surveys  becoming available, however, g-g
lensing  has now  been detected  with very  high significance, and  as
function  of   various  properties  of   the  lensing  galaxies  (e.g.
Griffiths \etal  1996; Hudson \etal  1998; McKay \etal 2001;  Guzik \&
Seljak 2002;  Hoekstra \etal 2003, 2004;  Sheldon \etal 2004, 2007a,b;
Mandelbaum \etal 2006; Heymans \etal 2006; Johnston \etal 2007; Parker
\etal  2007; Mandelbaum,  Seljak  \&  Hirata 2008).   Unfortunately, a
proper interpretation  of  these data  in terms  of  the  link between
galaxies and dark matter haloes has been hampered by the fact that the
lensing signal can typically only be detected when stacking the signal
of many  lenses.  Since not all  lenses  reside in haloes of  the same
mass, the resulting   signal is a non-trivial  average  of the lensing
signal due to  haloes of different masses.   Most studies to date have
assumed that the relation between the  luminosity of a lens galaxy and
the mass of its halo is given by a simple power-law relation with zero
scatter (see Limousin  \etal 2007 for  a detailed overview).  However,
it   has become clear,   recently, that the   scatter in this relation
between light and mass can be  very substantial (More \etal 2008b, and
references therein). As shown by Tasitsiomi \etal (2004), this scatter
has a very  significant impact on the actual  lensing signal, and thus
has  to be  accounted  for  in  the  analysis.   In addition,  central
galaxies (those residing   at the center  of  a dark matter halo)  and
satellite galaxies (those orbiting around a central galaxy) contribute
very different lensing signals, even when they reside in haloes of the
same mass  (e.g.   Natarajan, Kneib \&  Smail  2002; Yang  \etal 2006;
Limousin \etal 2007). This has to be  properly accounted for (see e.g.
Guzik  \& Seljak 2002), and  requires  knowledge of both the satellite
fractions and of the spatial number  density distribution of satellite
galaxies within their dark matter haloes.

Over  the years, numerous techniques have  been developed to interpret
galaxy-galaxy lensing measurements  (Schneider \&  Rix 1997; Natarajan
\&  Kneib  1997; Brainerd  \&  Wright 2002;  Guzik  \&   Seljak 2001).
Several  authors have also   used numerical simulations to investigate
the link between  g-g lensing and  the galaxy-dark matter  connnection
(e.g., Tasitsiomi \etal 2004; Limousin \etal 2005; Natarajan, De Lucia
\& Springel 2007;  Hayashi \& White 2007).   It has become  clear from
these studies that   g-g lensing  in  principle contains   a wealth of
information  regarding  the  mass distributions  around   galaxies; in
addition to  simply probing halo  masses, g-g  lensing also holds  the
potential  to measure the  shapes,  concentrations  and radii of  dark
matter haloes, and the first  observational results along these  lines
have already been obtained (Natarajan \etal 2002; Hoekstra \etal 2004;
Mandelbaum \etal 2006; Limousin \etal 2007; Mandelbaum \etal 2008).

In this paper we use an analytical model, similar to that developed by
Seljak (2000) and Guzik \&  Seljak (2001), to  predict the g-g lensing
signal as a function  of the luminosity of the  lenses starting from a
model  for the halo occupation statistics   that is constrained to fit
the abundances and  clustering  properties of  the  lens  galaxies.  A
comparison of these  predictions with the  data thus allows us to test
the mass-to-light ratios inferred from  the halo occupation model, and
ultimately to  constrain  cosmological parameters.  The  model assumes
that   haloes have  NFW  (Navarro,   Frenk  \&  White   1997)  density
distributions, and   that satellite  galaxies follow  a  radial number
density distribution that is unbiased with respect to the dark matter.
The occupation statistics are described via the conditional luminosity
function   (CLF;  see Yang \etal 2003),    which specifies the average
number of galaxies of given luminosity that reside in  a halo of given
mass. This CLF is ideally  suited to model g-g  lensing, as it  allows
one  to properly  account  for  the  scatter  in the relation  between
luminosity and   halo mass,  and  to  split the  galaxy  population in
centrals and satellites.  We   demonstrate how these different  galaxy
populations contribute  to the lensing  signal in different luminosity
bins, and  show    that    uncertainties related to     the   expected
concentrations of   dark matter haloes  and the  radial number density
distributions  of satellite galaxies  do not have a significant impact
on our results.

Assuming a flat   $\Lambda$CDM cosmology with parameters  supported by
the third   year data release  of  the  Wilkinson Microwave Anisotropy
Probe (WMAP, see Spergel \etal 2007), we  obtain a CLF that accurately
fits the abundances and clustering properties of SDSS galaxies.  Using
our analytical model, we   show that this   same CLF also   accurately
matches  the g-g lensing data obtained  from the SDSS  by Seljak \etal
(2005) and Mandelbaum \etal  (2006) {\it without any additional tuning
  of the  model parameters}. However, if we  repeat  the same exercise
for a cosmology with a matter density and power-spectrum normalization
that are slightly ($\sim$ 20 percent)  higher, the model that fits the
clustering data can no longer simultaneously fit the g-g lensing data.
This confirms that  a joint analysis of  clustering and g-g lensing is
an extremely promising method to constrain cosmological parameters. In
a companion  paper  (Li \etal  2008), we  use  the  SDSS galaxy  group
catalogue  of Yang \etal (2007)   to  predict the g-g lensing  signal,
which  we compare to data  from the SDSS.    Although, Li \etal obtain
their halo occupation statistics from a galaxy group catalogue, rather
than  from the galaxy clustering properties,  they obtain very similar
results.

The present  paper is  organized as follows.  We review  the necessary
formalism of  g-g lensing  in \S~\ref{sec:gglensing}, with  a detailed
description of the model used to interpret the g-g lensing signal. The
CLF, used to describe the  connection between galaxies and dark matter
haloes,  is  introduced in  \S~\ref{sec:CLF}.  The  properties of  the
predicted  g-g lensing signal  are illustrated  in \S~\ref{sec:predGG}
together with  a comparison  between theoretical predictions  and SDSS
data.  A  detailed analysis of  the assumptions entering the  model is
presented  in  \S~\ref{sec:moddep}.    Conclusions  are  presented  in
\S~\ref{sec:conclusions}.
 
\section{The Halo-Model Description of Galaxy-Galaxy Lensing}
\label{sec:gglensing}

Galaxy-galaxy  lensing  measures  the  tangential  shear  distortions,
$\gamma_{\rm  t}$, in  the  shapes of  background galaxies  (hereafter
sources) induced  by the mass distribution  around foreground galaxies
(hereafter lenses).   Since the tangential shear distortions  due to a
typical  lens  galaxy  (and  its  associated  dark  matter  halo)  are
extremely small, and since  background sources have non-zero intrinsic
ellipticities,    measuring   $\gamma_{\rm    t}$    with   sufficient
signal-to-noise requires large  numbers of background galaxies. Except
for extremely deep  surveys behind clusters of galaxies,  which have a
large  surface  area, the  number  density  of  background sources  is
insufficient  for a  reliable measurement  of $\gamma_{\rm  t}$ around
individual lenses.   In practice, this problem can  be circumvented by
stacking  many  lenses according  to  some  observable property.   For
example,  Mandelbaum et  al.  (2006)  measured $\gamma_{\rm  t}$  as a
function of the transverse comoving distance $R$ by stacking thousands
of lenses in a given  luminosity bin $[L_1,L_2]$.  The resulting shear
$\gamma_{\rm   t}(R|L_1,L_2)$    holds   information   regarding   the
characteristic mass  of the haloes that host  galaxies with luminosity
$L_1  \leq  L \leq  L_2$,  and  hence can  be  used  to constrain  the
galaxy-dark matter connection.

The tangential shear as a  function of the projected radius $R$ around
the lenses  is related  to the excess  surface density  (ESD) profile,
$\Delta \Sigma (R)$, according to
\begin{equation}\label{eq:shear} 
\Delta \Sigma (R) = \overline{\Sigma}(<R) - \Sigma(R) = 
\gamma_{\rm t}(R) \Sigma_{\rm crit} \,.
\end{equation}
where   $\Sigma(R)$    is   the   projected    surface   density   and
$\overline{\Sigma}(<R)$ is its average inside $R$,
\begin{equation}\label{eq:averageSigma}
\bar{\Sigma}(<R)  = \frac{2}{R^{2}}\int_{0}^{R}\Sigma(R') R' dR'\,,
\end{equation}
(Miralda-Escud\'e 1991;  Sheldon \etal 2004).   The so-called critical
surface density, $\Sigma_{\rm crit}$, is a geometrical parameter given
by
\begin{equation}\label{eq:sigcrit}
\Sigma_{\rm crit} = \frac{c^2}{4\pi G} 
\frac{\omega_{\rm S}}{\omega_{\rm L} \omega_{\rm LS}(1+z_{L})},
\end{equation}
with  $\omega_{\rm S}$,  $\omega_{\rm  L}$ and  $\omega_{\rm LS}$  the
comoving  distances to  the  source,  the lens  and  between the  two,
respectively, and with $z_{L}$ the redshift of the lens.

\begin{figure}
\centerline{\psfig{figure=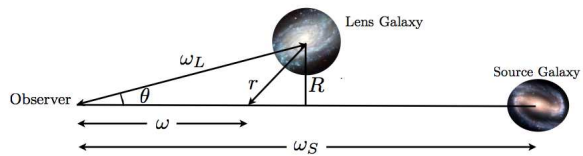,width=0.95\hssize}}
\caption{Illustration of the geometry between source, lens and
  observer}
\label{fig:sketch}
\end{figure}

The  projected surface density  is related  to the  galaxy-dark matter
cross correlation, $\xi_{\rm g,dm}(r)$, according to
\begin{equation}\label{eq:Sigma}
\Sigma(R) = \overline{\rho}\int_{0}^{\omega_\rms} 
\left[1+\xi_{\rm g,dm}(r)\right] \, \rmd \omega \, ,
\end{equation}
where  $\overline{\rho}$  is the  average  density  of  matter in  the
Universe and the integral is along the line of sight with $\omega$ the
comoving  distance from the  observer. The  three-dimensional comoving
distance $r$ is related to  $\omega$ through $r^2 = \omega_{\rm L}^2 +
\omega^2   -    2   \omega_{\rm   L}   \omega    \cos   \theta$   (see
Fig.~\ref{fig:sketch} for an illustration  of the geometry).  
Since $\xi_{\rm  g,dm}(r)$ goes  to zero in  the limit  $r \rightarrow
\infty$, and since  in practice $\theta$ is small,  we can approximate
Eq.\,(\ref{eq:Sigma}) using
\begin{equation}\label{eq:Sigma_approx}
\Sigma(R) = 2\overline{\rho}\int_{R}^{\infty} 
\left[1+\xi_{\rm g,dm}(r)\right] \, { r \rmd r \over \sqrt{r^2 - R^2}}\,,
\end{equation}
which is the expression we adopt throughout. 

The main goal of this paper is  to test the halo occupation statistics
inferred from galaxy  clustering  data with  g-g lensing  data. As is
evident from the above equations, the lensing signal $\Delta\Sigma(R)$
is completely specified by  the galaxy dark matter cross  correlation,
which,  as  we demonstrate below, can  be  computed from a  given halo
occupation model.  For computational convenience,  we will be  working
in Fourier space, where the related quantity is the galaxy-dark matter
cross power spectrum
\begin{equation}\label{Pgdmk}
P_{\rm g,dm}(k) = 4 \pi \int_{0}^{\infty} \xi_{\rm g,dm}(r)
{\sin(kr)\over kr} \, r^2 \, {\rm d}r\,.
\end{equation}
In order to  compute this power spectrum, we  follow Seljak (2000) and
Guzik \& Seljak  (2001), and adopt the halo  model, according to which
all  dark matter  is partitioned  over  dark matter  haloes (see  also
Mandelbaum \etal 2005a).  As usual in the halo model, it is convenient
to  split $P_{\rm  g,dm}(k)$  into  two terms;  a  1-halo term,  which
describes the  cross correlation between galaxies and  the dark matter
particles that reside in the same  halo, and a 2-halo term, where each
galaxy is cross  correlated with the dark matter  in all haloes except
for the  one that  hosts the galaxy  in question.  The  computation of
these two terms has to account for two important complications.  First
of all,  because of  the stacking procedure  used in order  to achieve
sufficient signal-to-noise,  the ESD contains signal  from haloes with
different masses.  A proper  estimate of $P_{\rm g,dm}(k)$, therefore,
requires  the full  probability distribution  that a  galaxy  with the
stacking property  used (in  this case luminosity)  resides in  a dark
matter halo  of mass $M$.  Secondly, central  galaxies (those residing
at the  center of  a dark matter  halo) and satellite  galaxies (those
orbiting around  a central  galaxy) contribute very  different lensing
signals, even when they reside in  haloes of the same mass (e.g., Yang
\etal  2006). This  has to  be  properly accounted  for, and  requires
knowledge of  both the satellite  fractions and of the  spatial number
density distribution  of satellite  galaxies within their  dark matter
haloes.  Based on these  considerations, we split $P_{\rm g,dm}(k)$ in
four terms:
\begin{eqnarray}\label{Pkfour}
P_{\rm g,dm}(k) 
&=& f_\rmc \left[P^{1\rm h,c}_{\rm g,dm}(k) + 
                 P^{2\rm h,c}_{\rm g,dm}(k)\right] \\ \nonumber
&+& f_\rms \left[P^{1\rm h,s}_{\rm g,dm}(k) + 
                 P^{2\rm h,s}_{\rm g,dm}(k)\right]\, ,
\end{eqnarray}
where `c' and `s'  stand for `central' and  `satellite', respectively.
The reason for explicitely writing the central and satellite fractions
($f_\rmc$ and $f_\rms = 1-f_\rmc$, respectively) in the above equation
will become apparent  below, in which we  describe each of these  four
terms in turn.

\subsection{The 1-halo term}
\label{sec:onehterm}

The one-halo  central term  of the power  spectrum describes  the dark
matter  distribution inside  haloes hosting  central galaxies.   For a
single,  central  lensing  galaxy,  it  simply  reflects  the  Fourier
transform of the overdensity of the dark matter halo in which the lens
resides:
\begin{equation}\label{eq:P1hcsingle}
P^{1\rm h,c}_{\rm g,dm}(k) = 
\frac{M}{\overline{\rho}} \, u_{\rm dm}(k|M)\,,
\end{equation}
where  $u_{\rm dm}(k|M)$ is  the normalized  Fourier transform  of the
mass density profile, $\rho(r|M)$:
\begin{equation}\label{eq:ukm}
u_{\rm dm}(k|M) = 4 \pi \int_{0}^{r_{180}} \frac{\rho(r|M)}{M} 
\frac{\sin(kr)}{kr} \, r^2 \, \rmd r\,.
\end{equation}
with  $r_{180}$ the  radius of  the halo  (see \S\ref{sec:ingredients}
below).  However,  because the lensing signal is  measured by stacking
galaxies with luminosities in the range $[L_1,L_2]$, we have that
\begin{equation}\label{eq:P1hc}
P^{1\rm h,c}_{\rm g,dm}(k) = \frac{1}{\overline{\rho}} 
\int_0^{\infty} \calP_\rmc(M|L_1,L_2) \, M \, u_{\rm dm}(k|M) \, \rmd M\,,
\end{equation}
where $\calP_\rmc(M|L_1,L_2)$ is the probability that a central galaxy
with luminosity $L_1  \leq L \leq L_2$ resides  in a halo of mass $M$.
This probability  function  reflects the  halo  occupation statistics,
and, using Bayes' theorem, can be written as
\begin{equation}\label{eq:prob1hc}
\calP_\rmc(M|L_1,L_2) \, \rmd M = 
\frac{\langle N_\rmc \rangle_{M}(L_1,L_2) \, n(M)}
{\overline{n}_\rmc(L_1,L_2)} \, \rmd M\,.
\end{equation}
Here $\langle  N_\rmc \rangle_{M}(L_1,L_2)$  is the average  number of
central  galaxies with  luminosities in  the range  $[L_1,  L_2]$ that
reside in a halo of mass $M$, $n(M)$ is the halo mass function and
\begin{equation}
\overline{n}_\rmc(L_1,L_2) = 
\int_0^{\infty} \langle N_\rmc \rangle_{M}(L_1,L_2) \, n(M) \, \rmd M\,,
\end{equation}
is  the comoving  number  density  of central  galaxies  in the  given
luminosity range.

Combining  (\ref{eq:P1hc}) and (\ref{eq:prob1hc}),  the first  term of
the galaxy-dark matter power spectrum can be written as
\begin{align}\label{eq:fcP1hc}
f_{\rm c}P^{1\rm h,c}_{\rm g,dm}(k) &=  \frac{1}{\bar{n}_{\rm tot} 
\bar{\rho}}   \nonumber \\
& \int \langle N_{\rm c} \rangle_{M}(L_{1},L_{2})\, M \, u_{\rm
  dm}(k|M) \, n(M) \,  {\rm d}M   
\end{align}
where $\overline{n}_{\rm  tot} = \overline{n}_\rmc(L_1,L_2)/f_\rmc$ is
the total  number density of  all galaxies (centrals  plus satellites)
with luminosities  in the range $[L_1,L_2]$.  Note  that, for brevity,
we don't  explicitely write the luminosity dependence  of $f_\rmc$ and
$\overline{n}_{\rm    tot}$,    but     it    is    understood    that
$f_\rmc=f_\rmc(L_1,L_2)$ and $\overline{n}_{\rm tot}=\overline{n}_{\rm
  tot}(L_1,L_2)$.

The  1-halo satellite  term is  similar  to the  1-halo central  term,
except  for the  fact that  satellite galaxies  do not  reside  at the
center  of  their  dark  matter  halo, but  follow  a  number  density
distribution $n_\rms(r|M)$.   Consequently, the 1-halo  lensing signal
due to satellite galaxies involves a convolution of $n_\rms(r|M)$ with
the mass  density profile $\rho(r|M)$ of  the host halo  in which they
reside.  Using  that in Fourier  space a convolution corresponds  to a
simple multiplication, we obtain:
\begin{equation}\label{eq:fsP1hs}
P^{1\rm h,s}_{\rm g,dm}(k) = 
\frac{1}{\overline{\rho}}\int_0^{\infty} \calP_\rms (M|L_1,L_2) 
\, u_\rms(k|M) \, M \, u_{\rm dm}(k|M) \, \rmd M\,,
\end{equation}
with 
\begin{equation}
u_\rms(k|M) = 4\pi \int_{0}^{r_{180}}
\frac{n_\rms(r|M)}{\langle N_\rms \rangle_M(L_1,L_2)} \frac{\sin(kr)}{kr} 
\, r^2 \, \rmd r
\end{equation}
the Fourier  transform of $n_\rms(r|M)$ normalized  by $\langle N_\rms
\rangle_M(L_1,L_2)$  which is  the average  number of  satellites with
$L_1 \leq L  \leq L_2$ that reside  in a halo of mass  $M$.  We assume
that  there is no  luminosity segregation  amongst the  satellites, so
that they  all follow  the same radial  profile, independent  of their
luminosity.   We write the  probability that  a satellite  galaxy with
$L_1 \leq L \leq L_2$ resides in a halo of mass $M$ as
\begin{equation}
\calP_\rms(M|L_1,L_2) \, \rmd M= 
\frac{\langle N_\rms \rangle_{M}(L_1,L_2) \, n(M)}
{\overline{n}_{s}(L_1,L_2)} \rmd M\ ,
\end{equation}
with 
\begin{equation}
\overline{n}_\rms(L_1,L_2) = \int_0^{\infty} 
\langle N_\rms\rangle_{M}(L_1,L_2) \, n(M) \, \rmd M\, ,
\end{equation}
the comoving number density of satellite galaxies with luminosities in
the range $[L_1,L_2]$.  The 1-halo  satellite term can thus be written
as
\begin{align}
f_{\rm s} & P^{1\rm h,s}_{\rm g,dm}(k)  =  \frac{1}{\bar{n}_{\rm tot} \bar{\rho}}  \nonumber \\
&  \int \langle N_{\rm s} \rangle_{M}(L_{1},L_{2}) \, u_{\rm s}(k|M) \, M \, u_{\rm dm}(k|M) \, n(M) \, {\rm d}M \ \ .
\end{align}
where we  have used  that $\overline{n}_{\rm  tot} =   \overline{n}_{\rm
  s}(L_1,L_2)/f_{\rm  s}$. Note that $f_\rms=f_\rms(L_1,L_2)$.

\subsection{The 2-halo term}
\label{sec:twohterm}

The  2-halo  term of  the  power  spectrum  describes the  correlation
between  galaxies  and dark  matter  particles  belonging to  separate
haloes.   Within the  halo model,  this means  cross  correlating each
galaxy with all the dark matter haloes other than the one in which the
galaxy in question resides. Using the fact that dark matter haloes are
a biased tracer of the dark matter mass distribution, the contribution
to the 2-halo term due to central galaxies can be written as
\begin{align}
P^{2\rm h,c}_{\rm g,dm}(k) = &
\frac{P^{\rm NL}_{\rm dm}(k)}{\overline{\rho}} 
\int_0^{\infty}  \calP_\rmc(M|L_1,L_2) \, b(M) \, \rmd M \ \ \ \ \ \ \ \ \ \ \ \ \nonumber \\
& \int_0^{\infty}  M' \, u_{\rm dm}(k|M') \, b(M') \, n(M') \, \rmd M'\,.
\end{align}
where  $P^{\rm NL}_{\rm dm}(k)$  and $b(M)$  are the  non-linear power
spectrum of the dark matter  and the halo bias function, respectively. 
The first integral reflects  the contribution of the central galaxies,
while  the second  integral describes  the dark  matter  density field
partitioned over haloes.  Using (\ref{eq:prob1hc}) we obtain
\begin{align}\label{eq:fcP2hc}
f_\rmc P^{2\rm h,c}_{\rm g,dm}(k) = &
 \frac{P^{\rm NL}_{\rm dm}(k)}{\overline{n}_{\rm tot}\overline{\rho}} 
 \int_0^{\infty} \langle N_\rmc \rangle_{M}(L_1,L_2) \, b(M) \, n(M) \, \rmd M  
 \nonumber \\
& \int_0^{\infty}  M' \, u_{\rm dm}(k|M') \, b(M') \, n(M') \, \rmd M'\,.
\end{align}
Similarly, the satellite part of the 2-halo term is given by
\begin{align}\label{eq:fsP2hs}
f_\rms  P^{2\rm h,s}_{\rm g,dm} & (k) = 
\frac{P^{\rm NL}_{\rm dm}(k)}{\overline{n}_{\rm tot}\overline{\rho}} 
\int_0^{\infty} M' \, u_{\rm dm}(k|M') \, b(M') \, n(M') \, \rmd M' \nonumber\\
& \int_0^{\infty} \langle N_\rms \rangle_{M}(L_1,L_2) \, u_\rms(k|M) 
\, b(M) \, n(M) \, \rmd M  \, .
\end{align}
where the second integral now accounts for the number density
distribution of satellite galaxies in haloes of mass $M$.

Note  that equations  (\ref{eq:fcP2hc})  and (\ref{eq:fsP2hs})  ignore
\textit{halo exclusion}, i.e. the fact that, in the halo model, haloes
can not overlap. In the  Appendix, we present an approximate method to
take halo exclusion into account. Far from being a detailed treatment,
the suggested  procedure accounts only  for the most  relevant effect,
i.e. the exclusion of dark  matter particles residing in the same host
halo of  central galaxies (see  Appendix for further  details). Unless
stated  otherwise, all  the  results shown  throughout  the paper  are
obtained applying halo exclusion as modelled in the Appendix.

In  addition, a  technical, as  well  as conceptual,  issue arises  in
calculating the 2-halo terms introduced in equations (\ref{eq:fcP2hc})
and  (\ref{eq:fsP2hs}).  Let  us rewrite  these two  equations  in the
following compact form:
\begin{align}
P^{2\rm h,c}_{\rm g,dm}(k) = &
P^{\rm NL}_{\rm dm}(k) \,  {\cal I}_{N_{\rm c}} \, {\cal I}_{M}(k) 
\nonumber \\
P^{2\rm h,s}_{\rm g,dm}(k) = &
P^{\rm NL}_{\rm dm}(k) \,  {\cal I}_{N_{\rm s}}(k) \, {\cal I}_{M}(k) \ ,
\end{align}
where  ${\cal I}_{N_{\rm  c}}$, ${\cal  I}_{N_{\rm s}}(k)$  and ${\cal
  I}_{M}(k)$ are
\begin{align}\label{eq:INc}
 {\cal I}_{N_{\rm c}}  =  &
 \int_0^{\infty} \frac{\langle N_\rmc \rangle_{M}(L_1,L_2)}{\overline{n}_{\rm c}} \, b(M) \, n(M) \, \rmd M \ , \nonumber \\
 {\cal I}_{N_{\rm s}}(k)  =  &
 \int_0^{\infty} \frac{\langle N_\rms \rangle_{M}(L_1,L_2)}{\overline{n}_{\rm s}} \, u_{\rm s}(k|M) \, b(M) \, n(M) \, \rmd M \ , \nonumber \\
  {\cal I}_{M}(k)  =  &
 \int_0^{\infty} \frac{M}{\overline{\rho}} u_{\rm dm}(k|M) \, b(M) \, n(M) \, \rmd M \ .
\end{align}
The evaluation of these integrals is  somewhat tedious numerically, as
it requires  knowledge  of the halo mass   function and the halo  bias
function  over the entire mass   range $[0,\infty)$. Since these  have
only been tested against numerical simulations over a limited range of
halo    masses    ($10^{9}h^{-1}M_{\odot}           \lta    M     \lta
10^{15}h^{-1}M_{\odot}$),  it is also unclear   how accurate they are.
In  practice, though, these  problems can be  circumvented as follows.
First  of all, because  of the  exponential cut-off  in the halo  mass
function, it is  sufficiently accurate to  perform the integrations of
Eq.~(\ref{eq:INc})    only  up to  $M =     10^{16} h^{-1} M_{\odot}$.
Secondly, ${\cal I}_{N_{\rm c}}$ and ${\cal I}_{N_{\rm s}}(k)$ contain
the   halo     occupation     statistics,   $\langle      N_{\rm    c}
\rangle_{M}(L_{1},L_{2})$      and      $\langle        N_{\rm      s}
\rangle_{M}(L_{1},L_{2})$,  respectively, which,  for all luminosities
of interest in this paper, are equal to zero for $M \lta 10^{9} h^{-1}
M_{\odot}$.  Therefore, ${\cal I}_{N_{\rm  c}}$ and ${\cal  I}_{N_{\rm
    s}}(k)$ can  be computed accurately  by only integrating  over the
mass range  $[10^{9}-10^{16}]  h^{-1} M_{\odot}$.   Unfortunately, the
integrand of ${\cal I}_{M}(k)$  does not become negligibly small below
a  given halo mass.   However, in this case we   can use  the approach
introduced by Yoo \etal (2006): we write ${\cal  I}_{M}(k)$ as the sum
of  two    terms,  ${\cal I}_{M}(k)    =   {\cal   I}_{M_{1}}(k)+{\cal
  I}_{M_{2}}(k)$, where:
\begin{eqnarray}
{\cal I}_{M_{1}}(k) & = & \int_{0}^{M_{\rm min}} \frac{M}{\bar{\rho}} \ u_{\rm dm}(k|M) \ b(M) \ n(M) \ {\rm d}M \ , \nonumber \\
{\cal I}_{M_{2}}(k) & = & \int_{M_{\rm min}}^{\infty}  \frac{M}{\bar{\rho}} \ u_{\rm dm}(k|M) \ b(M) \ n(M) \ {\rm d}M \ .
\end{eqnarray}
Following Yoo \etal (2006), we  use the fact that $u_{\rm dm}(k|M) =1$
over the relevant  range of $k$ as long as $M$  is sufficiently small. 
This allows us to write
\begin{eqnarray}
{\cal I}_{M_{1}}(k) & \simeq & \int_{0}^{M_{\rm min}}  
\frac{M}{\bar{\rho}}  \ b(M) \ n(M) \ {\rm d}M \nonumber \\
& = & 1 - \int_{M_{\rm min}}^{\infty}  \frac{M}{\bar{\rho}} \ b(M) \ n(M) \ {\rm d}M  \ .
\end{eqnarray}
where the last equality follows from the fact that the distribution of
matter  is by definition  unbiased with  respect to  itself.  Detailed
tests  have  shown  that   this  procedure  yields  results  that  are
sufficiently  accurate as  long as  $M_{\rm min}  \lta  10^{10} h^{-1}
M_{\odot}$. Throughout we adopt $M_{\rm min} = 10^{9}h^{-1}M_{\odot}$.

\subsection{Model Ingredients}
\label{sec:ingredients}

The computation  of the galaxy-dark  matter cross correlation  (or its
power spectrum) as outlined  in the previous subsections, requires the
following ingredients:
\begin{itemize}

\item The halo mass function, $n(M)$, specifying the comoving number
  density of dark matter haloes of mass $M$.
  
\item The  halo bias function,  $b(M)$, which describes how  haloes of
  mass  $M$  are  biased  with  respect  to  the  overall  dark  matter
  distribution.
  
\item The  non-linear power spectrum of the  dark matter distribution,
  $P^{\rm NL}_{\rm dm}(k)$.

\item The mass density distribution of dark matter haloes, $\rho(r|M)$.
  
\item The  number density distribution  of satellite galaxies  in dark
  matter haloes, $n_\rms(r|M)$.
  
\item  The  halo  occupation  statistics  for  central  and  satellite
  galaxies,  as parameterized  by  $\langle N_{\rm  c} \rangle_M$  and
  $\langle N_\rms \rangle_M$, respectively.

\end{itemize}

All these ingredients depend on  cosmology.  In this paper we consider
two  flat $\Lambda$CDM cosmologies.   The first  has a  matter density
$\Omega_{\rm  m}=0.238$,   a  baryonic  matter   density  $\Omega_{\rm
  b}=0.041$,  a   Hubble  parameter  $h=H_0/(100   \kmsmpc)=0.734$,  a
power-law initial  power spectrum with spectral index  $n=0.951$ and a
normalization  $\sigma_8=0.744$.    These  are  the   parameters  that
best-fit the 3-year data release of the Wilkinson Microwave Anisotropy
Probe (WMAP, see Spergel \etal 2007), and we will refer to this set of
cosmological parameters as the  WMAP3 cosmology.  The second cosmology
has $\Omega_{\rm m}=0.3$,  $\Omega_{\rm b}=0.04$, $h=0.7$, $n=1.0$ and
$\sigma_8=0.9$.  With strong support  from the first year data release
of  the  WMAP  mission  (see  Spergel  \etal  2003),  this  choice  of
parameters  has been  considered in  many previous  studies.   In what
follows we  will refer to this  second set of parameters  as the WMAP1
cosmology.  For clarity, the parameters of both cosmologies are listed
in Table~1.
\begin{table}
\label{tab:cospar}
\caption{Cosmological Parameters}
\begin{tabular}{ccccccc}
\hline
 & $\Omega_{\rm m}$ & $\Omega_{\Lambda}$ & $\Omega_{\rm b}$ & $h$ & $n$ & $\sigma_{8}$ \\
\hline
\hline
WMAP3 & $0.238$ & $0.762$ & $0.041$ & $0.734$ & $0.951$ & $0.744$ \\
WMAP1 & $0.3$ & $0.7$ & $0.040$ & $0.7$ & $1.0$ & $0.9$ \\
\hline
\end{tabular}
\medskip
\begin{minipage}{\hsize}
\end{minipage}
\end{table}

We define dark matter haloes as spheres with an average overdensity of
$180$, with a mass given by
\begin{equation}\label{eq:mass_def}
M = \frac{4 \pi}{3} \, (180\overline{\rho}) \, r_{180}^3\,.
\end{equation}
Here $r_{180}$ is the radius of  the halo.  We assume that dark matter
haloes  follow  the  NFW   (Navarro,  Frenk  \&  White  1997)  density
distribution
\begin{equation}\label{NFW}
\rho(r) = 
\frac{\overline{\delta}\,\overline{\rho}}{(r/r_{*})(1+r/r_{*})^{2}}\,,
\end{equation}
where $r_{*}$ is a  characteristic radius and $\overline{\delta}$ is a
dimensionless amplitude  which can be  expressed in terms of  the halo
concentration parameter $c_{\rm dm} \equiv r_{180}/r_{*}$ as
\begin{equation}\label{overdensity}
\overline{\delta} = {180 \over 3} \, {c_{\rm dm}^{3} \over 
{\rm ln}(1+c_{\rm dm}) - c_{\rm dm}/(1+c_{\rm dm})}\,.
\end{equation}
Numerical simulations  show that $c_{\rm dm}$ is  correlated with halo
mass,  and  we use  the  relations  given  by Macci\`o  \etal  (2007),
converted to our definition of halo mass.

For the halo mass function, $n(M)$, and halo bias function, $b(M)$, we
use the functional  forms suggested by  Warren \etal (2006) and Tinker
\etal  (2005), respectively,  which   have been  shown to   be in good
agreement with numerical   simulations.  The linear power spectrum  of
density   perturbations is computed  using   the transfer function  of
Eisenstein \&  Hu  (1998), which  properly accounts  for the  baryons,
while the evolved, non-linear  power   spectrum of the dark    matter,
$P^{\rm NL}_{\rm  dm}(k)$, is computed  using  the fitting  formula of
Smith \etal (2003).

For  the  number density  distribution of  the  satellite galaxies, we
assume a generalized NFW profile (e.g., van den Bosch \etal 2004):
\begin{equation}\label{eq:generalisedNFW}
n_\rms(r|M) \propto \left({r\over \calR r_*} \right)^{-\alpha}
\left(1 + {r\over \calR r_*} \right)^{\alpha-3}\,,
\end{equation}
which  scales  as $n_\rms  \propto  r^{-\alpha}$  and $n_\rms  \propto
r^{-3}$ at small  and large radii, respectively.  Similar  to the dark
matter  mass distribution,  $n_{\rm  s}(r|M)$ has  an effective  scale
radius  $\calR r_*$, and  can be  parameterized via  the concentration
parameter $c_{\rm  s} = c_{\rm dm}/\calR$. Observations  of the number
density distribution of satellite galaxies in clusters and groups seem
to  suggest  that $n_\rms(r|M)$  is  in  good  agreement with  an  NFW
profile, for  which $\alpha=1$ (e.g.,  Beers \& Tonry  1986; Carlberg,
Yee  \&  Ellingson 1997a;  van  der Marel  \etal  2000;  Lin, Mohr  \&
Stanford  2004; van  den  Bosch \etal  2005a).   Several studies  have
suggested,  however, that  the satellite  galaxies are  less centrally
concentrated than the dark matter, corresponding to $\calR > 1$ (e.g.,
Yang  \etal 2005;  Chen 2007;  More \etal  2008b).  For  our fiducial
model we  adopt $\alpha=\calR=1$, for  which $u_{\rm s}(k|M)  = u_{\rm
  dm}(k|M)$ (i.e.,  satellite galaxies follow the  same number density
distribution as the  dark matter particles).  In \S\ref{sec:impactSat}
we examine how the results depend on $\alpha$ and $\calR$.

The  final ingredient is a model  for the  halo occupation statistics.
In their attempt  to model the g-g  lensing  signal obtained  from the
SDSS, Seljak  \etal  (2005)  and  Mandelbaum  \etal (2006)   made  the
oversimplified assumption  of a deterministic relation between central
galaxy luminosity and host halo mass. In particular, they used
\begin{equation}
\langle N_\rmc \rangle_{M}(L_1,L_2)=
\left\{
\begin{array}{cl}
1 & {\rm if} \, M = \widetilde{M}(L_1,L_2) \\
0 & {\rm otherwise}
\end{array}
 \right.
\end{equation}
where $\widetilde{M}(L_1,L_2)$ is the  `characteristic' mass of a halo
that  hosts a central galaxy  with $L_1 \leq  L \leq L_2$.  However, a
realistic  relation between central galaxy   luminosity and host  halo
mass will have  some  scatter.  As  demonstrated  by Tasitsiomi  \etal
(2004), this scatter  can have an  important impact on the g-g lensing
signal (see  also \S\ref{sec:impactScatter} below).  For the satellite
galaxies,  Seljak \etal (2005) and  Mandelbaum  \etal (2006) adopted a
simple double power-law relation of the form
\begin{equation}
\langle N_\rms \rangle_{M}(L_1,L_2) \propto
\left\{
\begin{array}{cl}
M   & {\rm if} \, M \geq 3\widetilde{M}(L_1,L_2) \\
M^2 & {\rm otherwise}
\end{array}
 \right.
\end{equation}

In this paper we improve upon  the analysis by Seljak \etal (2005) and
Mandelbaum \etal (2006)  by using a more realistic  model for the halo
occupation statistics. Furthermore, rather  than fitting the model to
the  lensing  data,  we  constrain  the  occupation  statistics  using
clustering  data from  the SDSS  combined  with a  large galaxy  group
catalogue.  Subsequently  we  use that model  to predict  the g-g
lensing signal which we compare  to g-g lensing data obtained from the
SDSS.

As  a final  remark,  we  emphasise that  different  quantities, e.g.  
$n(M)$, $b(M)$, and $P^{\rm NL}_{\rm dm}(k)$, depend on redshift, $z$,
even though we have not made this explicit in the equations.

\section{Conditional Luminosity Function}
\label{sec:CLF}

\subsection{Model description}
\label{sec:CLFmodel}

In order  to specify the halo  occupation statistics, we  use the CLF,
$\Phi(L|M)\rmd L$, which specifies the average number of galaxies with
luminosities in  the range $L \pm \rmd  L/2$ that reside in  a halo of
mass $M$  (Yang, Mo  \& van den  Bosch 2003;  van den Bosch,  Yang, Mo
2003). Following  Cooray \& Milosavljevi\'c (2005)  and Cooray (2006),
we write the CLF as
\begin{equation}
\Phi(L|M) = \Phi_\rmc(L|M) + \Phi_\rms(L|M)\, ,
\end{equation}
where $\Phi_\rmc(L|M)$ and   $\Phi_\rms(L|M)$  represent  central  and
satellite galaxies, respectively.  The occupation numbers required for
the  computation of  the   galaxy-dark matter cross correlation   then
simply follow from
\begin{equation}\label{eq:Ns}
\langle N_{\rm x} \rangle_{M}(L_1,L_2) =
\int_{L_1}^{L_2} \Phi_{\rm x}(L|M) \rmd L\,.
\end{equation}
where   `x'  refers to either   `c'   (centrals) or `s'  (satellites).
Motivated by the results of Yang, Mo \& van den Bosch (2008; hereafter
YMB08), who  analyzed  the CLF obtained  from  the  SDSS  galaxy group
catalogue of  Yang \etal (2007),  we assume the  contribution from the
central galaxies to be a log-normal:
\begin{equation}\label{eq:phi_c}
\Phi_\rmc(L|M) = 
{1\over {\sqrt{2\pi} \, {\rm ln}(10) \, \sigma_\rmc \, L}} {\rm exp}
\left[- { {(\log L  -\log L_\rmc )^2 } \over 2\sigma_\rmc^2} \right]\,.
\end{equation}
Note  that  $\sigma_\rmc$ is  the  scatter  in  $\log L$  (of  central
galaxies)  at  a fixed  halo  mass.  Moreover,  $\log L_\rmc$  is,  by
definition, the expectation value  for the (10-based) logarithm of the
luminosity of the central galaxy, i.e.,
\begin{equation}
\log L_\rmc = \int_0^{\infty} \Phi_\rmc(L|M) \, \log L \, \rmd L\,.
\end{equation}
For the contribution  from the satellite galaxies  we adopt a modified
Schechter function:
\begin{equation}\label{eq:phi_s}
\Phi_\rms(L|M) = {\phi^*_\rms\over L^*_\rms} 
\left({L\over L^*_\rms}\right)^{\alpha_\rms}
{\rm exp} \left[- \left ({L\over L^*_\rms}\right )^2 \right].
\end{equation}
which decreases  faster than a Schechter  function at the  bright end. 
Note  that $L_\rmc$,  $\sigma_\rmc$, $\phi^*_\rms$,  $\alpha_\rms$ and
$L^*_\rms$  are  all   functions  of  the  halo  mass   $M$.   In  the
parametrization of these mass dependencies, we again are guided by the
results of  YMB08.  In particular,  for the luminosity of  the central
galaxies we adopt
\begin{equation}\label{eq:LcM}
L_\rmc(M) = L_0 {(M/M_1)^{\gamma_1} \over 
\left[1 + (M/M_1) \right]^{\gamma_1-\gamma_2}}\,,
\end{equation}
so that $L_\rmc \propto M^{\gamma_1}$ for $M \ll M_1$ and $L_c \propto
M^{\gamma_2}$ for  $M \gg  M_1$. Here $M_1$  is a  characteristic mass
scale, and $L_0 =  2^{\gamma_1-\gamma_2} L_c(M_1)$ is a normalization. 
Using  the SDSS  galaxy  group  catalogue, YMB08  found  that to  good
approximation
\begin{equation}
L^*_\rms(M)  = 0.562 L_\rmc(M)
\end{equation}
and  we adopt  this  parameterization throughout.   For the  faint-end
slope and normalization of $\Phi_\rms(L|M)$ we adopt
\begin{equation}\label{eq:alpha}
\alpha_\rms(M) = -2.0 + 
a_1\left( 1- {2 \over \pi}{\rm arctan}[a_2 \log(M/M_2)]\right)\,,
\end{equation}
and
\begin{equation}\label{eq:phi}
\log[\phi^*_\rms(M)] = b_0 + b_1 (\log M_{12}) + b_2 (\log M_{12})^2\,,
\end{equation}
with $M_{12}=M/(10^{12}  h^{-1}\Msun)$. This adds a total  of six free
parameters: $a_1$,  $a_2$, $b_0$, $b_1$, $b_2$  and the characteristic
halo mass $M_{2}$.   Neither of these functional forms  has a physical
motivation; they merely were  found to adequately describe the results
obtained by YMB08. Finally, for simplicity, and to limit the number of
free  parameters, we  assume that  $\sigma_\rmc(M) =\sigma_\rmc$  is a
constant. As shown in More \etal (2008b), this assumption is supported
by the kinematics of satellite galaxies in the SDSS.  Thus, altogether
the CLF has a total of eleven free parameters.
\begin{table}
\label{tab:wangdata}
\caption{Correlation lengths.}
\begin{tabular}{cccc}
   \hline
Sample & $^{0.1}M_r - 5 \log h$ & $<z>$ & $r_0$ \\
\hline\hline
V1 & $(-23.0,-21.5]$ & $0.173$ & $7.59 \pm 0.75$ \\
V2 & $(-21.5,-21.0]$ & $0.135$ & $6.11 \pm 0.33$ \\
V3 & $(-21.0,-20.5]$ & $0.109$ & $5.62 \pm 0.16$ \\
V4 & $(-20.5,-20.0]$ & $0.089$ & $5.38 \pm 0.16$ \\
V5 & $(-20.0,-19.0]$ & $0.058$ & $4.90 \pm 0.18$ \\
V6 & $(-19.0,-18.0]$ & $0.038$ & $4.17 \pm 0.23$ \\
\hline
\end{tabular}
\medskip
\begin{minipage}{\hssize}
  Galaxy-galaxy  clustering correlation lengths  of Wang  \etal (2007)
  used in this paper to constrain the CLF.  Column (1) lists the ID of
  each volume  limited sample, following  the notation of Wang  et~al. 
  (2007).  Columns  (2) and (3) indicate the  absolute magnitude range
  and the  mean redshift  of each sample,  while column (4)  lists the
  correlation  length plus its  standard deviation  (in $h^{-1}\Mpc$),
  obtained  by  fitting  a  power-law  to  the  projected  correlation
  function over the  radial range $[0.98, 9.6] h^{-1}  \Mpc$. See Wang
  \etal (2007) for details.
\end{minipage}
\end{table}
\begin{figure*}
\centerline{\psfig{figure=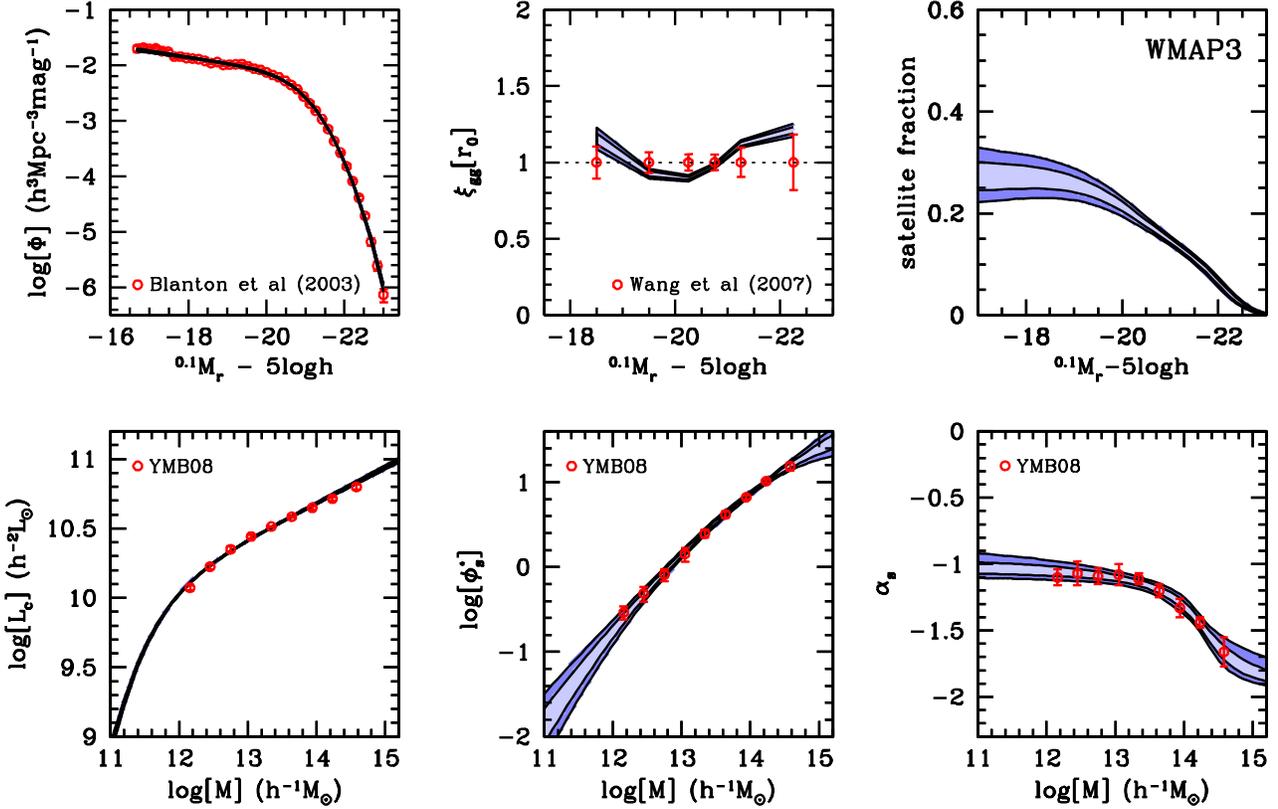,width=0.95\hdsize}}
\caption { \textit{Upper row, left and central panels}. The luminosity
  function  of galaxies and  the luminosity  dependence of  the galaxy
  correlation  length are  plotted.  Data  come from  the  analysis of
  Blanton  \etal  (2003a) and  Wang  \etal  2007.   The blue  contours
  indicate the  68 and 95  percent confidence level obtained  from the
  MCMC.   The  agreement  is  extremely accurate  for  the  luminosity
  function  whereas   is  satisfying  for  the   correlation  length.  
  \textit{Lower row, three panels}.  The additional information coming
  from  the group  catalogue of  YMB08  is plotted  together with  the
  corresponding 68  and 95 percent  confidence level derived  with the
  MCMC. In particular, the halo  mass dependence of the central galaxy
  luminosity,   the    satellite   conditional   luminosity   function
  normalization $\phi^{*}_{\rm  s}$ and the  the exponent $\alpha_{\rm
    s}$ are shown in the left, central and middle panel, respectively.
  \textit{Upper row,  right panel}.  The 68 and  95 percent confidence
  levels of the satellite  fraction, $f_{\rms}$, obtained from the CLF
  (see eq.~[\ref{fsatL}]).}
\label{fig:CLFres3}
\end{figure*}
\begin{figure*}
\centerline{\psfig{figure=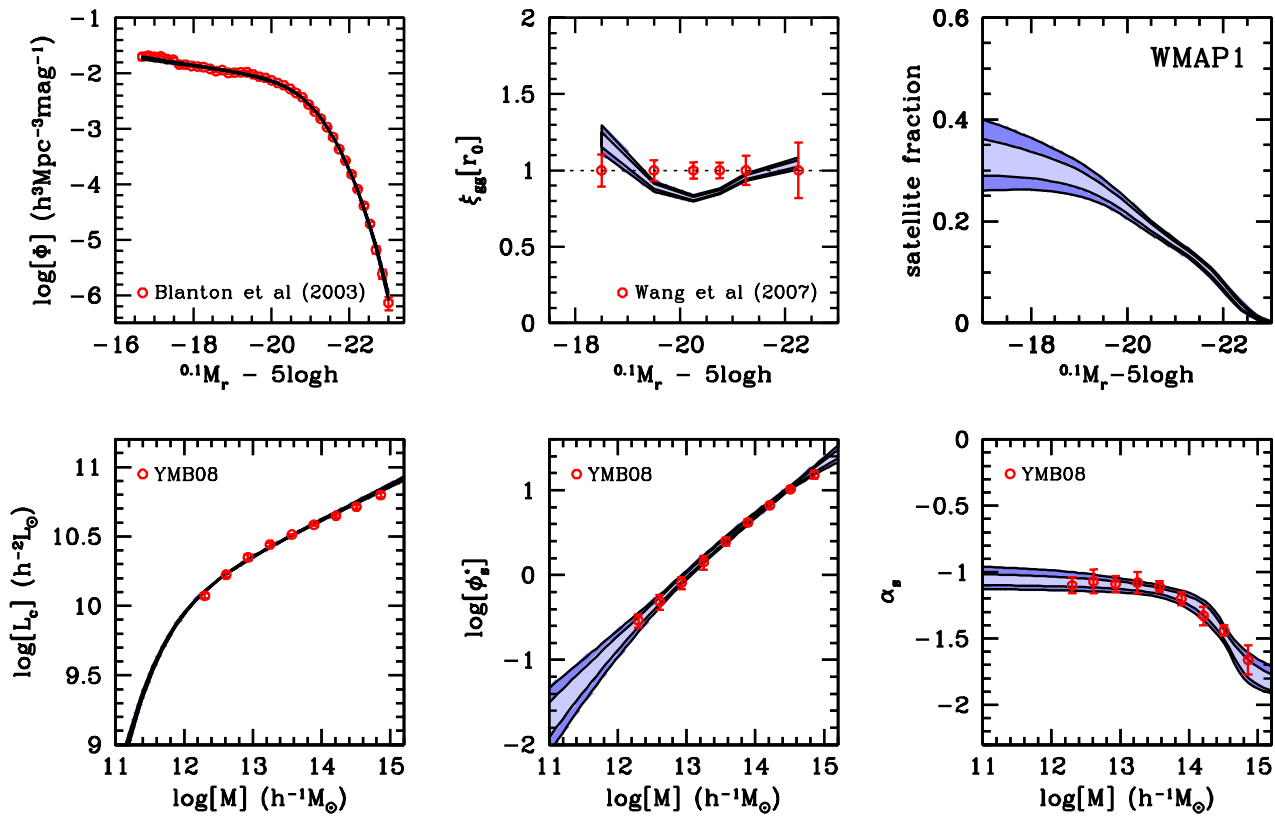,width=0.95\hdsize}}
\caption {Same as Fig.~\ref{fig:CLFres3} but for the WMAP1 cosmology.}
\label{fig:CLFres1}
\end{figure*}

Note that, with  the parametrization of  the CLF introduced above, the
halo occupation statistics can be rewritten as:
\begin{equation}
\langle N_{\rm c} \rangle_{M}(L_1,L_2) = 
\int_{L_1}^{L_2} \Phi_{\rm c}(L|M) \rmd L  =  
\frac{1}{2}\Big[  \textrm{erf} (x_{2}) - \textrm{erf}(x_{1}) \Big]
\end{equation}
\begin{eqnarray}
& & \langle N_{\rm s} \rangle_{M}(L_1,L_2) =
\int_{L_1}^{L_2} \Phi_{\rm s}(L|M) \rmd L  = \nonumber \\
& & \frac{\phi^{*}_{\rm s}}{2}
\Bigg\{ 
\Gamma
\left[\frac{\alpha_{\rm s}}{2} + \frac{1}{2}, \left( \frac{L_{1}}{L^{*}_{\rm s}} \right)^{2}\right]
-  
\Gamma 
\left[ \frac{\alpha_{\rm s}}{2} + \frac{1}{2}, \left( \frac{L_{2}}{L^{*}_{\rm s}} \right)^{2} \right]
\Bigg\} \, ,\nonumber \\
& & 
\end{eqnarray}
where erf$(x_{i})$ is the error  function calculated at $x_{i} =  \log
(L_{i}/L_{\rm  c})/(\sqrt{2}\sigma_{\rm c})$   with   $i =  1,2$   and
$\Gamma$ is the incomplete gamma function.

As shown in Yang \etal (2003)  and van den Bosch \etal (2003), the CLF
can  be  constrained  using   the  observed  luminosity function,  $\Phi(L)$,  and  the
galaxy-galaxy  correlation  lengths   as  a  function  of  luminosity,
$r_0(L)$.   Here we  use the  luminosity function (hereafter LF) of Blanton  \etal (2003a)  uniformly
sampled at  41 magnitudes covering  the range $-23.0  \leq {^{0.1}M}_r
-5\log  h  \leq -16.4$.   Here  ${^{0.1}M}_r$  indicates the  $r$-band
magnitude  K$+$E  corrected  to  $z=0.1$ following  the  procedure  of
Blanton  \etal (2003b).  For  the correlation  lengths as  function of
luminosity we  use the results obtained  by Wang \etal  (2007) for six
volume limited samples selected  from the SDSS DR4.  For completeness,
these  data  are  listed  in  Table~2.   Finally,  to  strengthen  our
constraints, and  to assure agreement  with the CLF obtained  from our
SDSS   group  catalogue,   we   use  the   constraints  on   $L_c(M)$,
$\alpha_s(M)$ and $\phi^*_s(M)$ obtained by YMB08.

For a given set of model parameters, we compute the LF using
\begin{equation}
\Phi(L) = \int_0^{\infty} \Phi(L|M) \, n(M) \, \rmd M\,.
\end{equation}
The galaxy-galaxy correlation  function for galaxies with luminosities
in the interval $[L_1,L_2]$ is computed using
\begin{equation}
\xi_{\rm gg}(r) = b_{\rm gal}^2(L_1,L_2) \, \zeta(r) \,
\xi_{\rm dm}^{\rm NL}(r)\,.
\end{equation}
Here $\xi_{\rm dm}^{\rm NL}(r)$ is the non-linear correlation function
of the dark matter, which is the Fourier transform of $P^{\rm NL}_{\rm
  dm}(k)$,
\begin{equation}
\zeta(r) = {
[1+1.17\xi_{\rm dm}^{\rm NL}(r)]^{1.49} \over 
[1+0.69\xi_{\rm dm}^{\rm NL}(r)]^{2.09}}\,,
\end{equation}
is the radial scale dependence of the bias as obtained by Tinker \etal
(2005), and $b_{\rm gal}(L_1,L_2)$ is  the bias of the galaxies, which
is related to the CLF according to
\begin{equation}
b_{\rm gal}(L_1,L_2) = {\int_0^{\infty} \langle N \rangle_M \, b(M) \,
  n(M) \, \rmd M \over \int_0^{\infty} \langle N \rangle_M \,
  n(M) \, \rmd M}\,,
\end{equation}
with 
\begin{eqnarray}
\langle N \rangle_M & = & \int_{L_1}^{L_2} \Phi(L|M) \rmd L\nonumber\\
 & = & \langle N_\rmc \rangle_M(L_1,L_2) + 
\langle N_\rms \rangle_M(L_1,L_2)\,.
\end{eqnarray}
the  average  number  of  galaxies  with  luminosities  in  the  range
$[L_1,L_2]$ that reside in a halo of mass $M$.

\begin{table*}
\label{tab:CLFpar}
\caption{Best-fit CLF parameters obtained from SDSS clustering analysis}
\begin{tabular}{ccccccccccccc}
\hline
Cosmology & $\log L_{0}$ & $\log M_{1}$ & $\gamma_{1}$ & $\gamma_{2}$ & 
$a_{1}$ & $a_{2}$ & $\log M_{2}$ & $b_{0}$& $b_{1}$& $b_{2}$ & 
$\sigma_{\rm c}$ & $\chi^2_{\rm red}$ \\
 (1) & (2) & (3) & (4) & (5) & (6) & (7) & (8) & (9) & (10) & (11) &
 (12) & (13) \\
\hline
\hline
WMAP3 & 9.935 & 11.07 & 3.273 & 0.255  & 0.501 & 2.106 & 14.28 &
-0.766 & 1.008 & -0.094 & 0.143 & 1.42 \\
WMAP1 & 9.915 & 11.16 & 3.336 & 0.248 & 0.484 & 2.888 & 14.54 & 
-0.854 & 0.906 & -0.062 & 0.140 & 1.70 \\ 
\hline
\end{tabular}
\medskip
\begin{minipage}{\hsize}
  The best-fit CLF parameters obtained  from the MCMC analysis for the
  WMAP3  and WMAP1  cosmologies  and the  value  of the  corresponding
  reduced $\chi^2$. Masses and  luminosities are in $h^{-1} \Msun$ and
  $h^{-2} \Lsun$, respectively.
\end{minipage}
\end{table*}

To determine the likelihood function  of our free parameters we follow
van  den Bosch  \etal  (2007)  and use  the  Monte-Carlo Markov  Chain
(hereafter  MCMC)  technique. The  goodness-of-fit  of  each model  is
judged using $\chi^2 = \chi^2_{\Phi} + \chi^2_{r_0} + \chi^2_{\rm GC}$
with
\begin{equation}
\label{chisqLF}
\chi^2_{\Phi} = \sum_{i=1}^{41}
\left[ {\Phi(L_i) - \hat{\Phi}(L_i) \over \Delta \hat{\Phi}(L_i)} \right]^2\,,
\end{equation}
\begin{equation}
\label{chisqr0}
\chi^2_{r_0} = \sum_{i=1}^{6}
\left[ {\xi_{\rm gg}(r_{0,i}) - 1 \over 
\Delta \hat{\xi}_{\rm gg}(r_{0,i})} \right]^2\,,
\end{equation}
and 
\begin{eqnarray}
\label{chisqr0}
\chi^2_{\rm GC} & = & \sum_{i=1}^{9}
\left[ {\log L_c(M_i) - \log\hat{L}_c(M_i) \over 
\Delta \log\hat{L}_c(M_i)} \right]^2 \nonumber \\
 & + & \sum_{i=1}^{9}
\left[ {\alpha_s(M_i) - \hat{\alpha}_s(M_i) \over 
\Delta \hat{\alpha}_s(M_i)} \right]^2 \nonumber \\
 & + & \sum_{i=1}^{9}
\left[ {\phi^*_s(M_i) - \hat{\phi}^*_s(M_i) \over 
\Delta \hat{\phi}^*_s(M_i)} \right]^2\,.
\end{eqnarray}
Here   $\hat{.}$ indicates an observed    quantity and the  subscripts
`$\Phi$', `$r_{0}$'   and `GC' refer to  the  luminosity function, the
galaxy-galaxy  correlation    length    and  the    group   catalogue,
respectively. Note that, by definition, $\hat{\xi}_{\rm gg}(r_{0,i}) =
1$.    Table~4 lists the best-fit parameters    obtained with the MCMC
technique for both  the WMAP1 and  WMAP3  cosmologies, as well  as the
corresponding   value  of $\chi^2_{\rm  red}=\chi^2/N_{\rm dof}$. Here
$N_{\rm dof} = 74-11=63$ is the number of degrees of freedom.

\subsection{Results}
\label{sec:CLFres}

Fig.~\ref{fig:CLFres3}  shows  the  results  obtained  for  the  WMAP3
cosmology.  In  each panel  the blue contours  indicate the 68  and 95
percent confidence levels obtained  from the MCMC. The upper left-hand
panel  shows that  the  CLF model  accurately  fits the  galaxy LF  of
Blanton \etal (2003a).  The fit to the correlation lengths as function
of  luminosity, shown  in the  upper middle  panel, is  less accurate,
although data and model typically  agree at the $1 \sigma$ level.  The
lower  panels of  Fig.~\ref{fig:CLFres3} show  the 68  and  95 percent
confidence   levels   on   $L_{\rm  c}(M)$,   $\phi^{*}_\rms(M)$   and
$\alpha_\rms(M)$, compared with the results obtained by YMB08 from the
SDSS  group catalogue  of Y07.   Since these  data have  been  used as
additional constraints, it should not  come as a big surprise that the
CLF is in good agreement  with these data.  We emphasise, though, that
it is  not trivial that  a single halo  occupation model can  be found
that can simultaneously  fit the LF, the luminosity  dependence of the
galaxy-galaxy correlation  functions, and the results  obtained from a
galaxy group catalogue.

Finally,  the upper right-hand  panel of  Fig.~\ref{fig:CLFres3} shows
the satellite fraction,
\begin{equation}\label{fsatL}
f_\rms(L) = {1 \over \Phi(L)} \,
\int_0^{\infty} \Phi_\rms(L|M) \, n(M) \, \rmd M\,.
\end{equation}
as function of luminosity.  This is found  to decrease from $\sim 0.27
\pm 0.03$ at  $^{0.1}M_r   - 5 \log h    = -17$ to virtually   zero at
$^{0.1}M_r - 5 \log h  = -23$. The fact   that the satellite  fraction
decreases with increasing  luminosity is in qualitative agreement with
previous studies  (Tinker \etal 2006; Mandelbaum  \etal 2006;  van den
Bosch \etal 2007).

We  have repeated  the  same  exercise for  the  WMAP1 cosmology.   As
evident from Fig.~\ref{fig:CLFres1}, for  this cosmology we can obtain
a CLF that fits the data  almost equally well (the reduced $\chi^2$ is
only slightly  higher than  for the WMAP3  cosmology; see  Table~3). 
Note that the group data (shown  in the lower panels) differ from that
in Fig.~\ref{fig:CLFres3}, even though the group catalogue is the same.
This owes to the fact that the halo mass assignments of the groups are
cosmology dependent  (see Y07  for details).  The  satellite fractions
inferred for  this cosmology, shown  in the upper right-hand  panel of
Fig.~\ref{fig:CLFres1}, are similar,  though slightly higher, than for
the WMAP3 cosmology,  in excellent agreement with van  den Bosch \etal
(2007).

\begin{figure}
\centerline{\psfig{figure=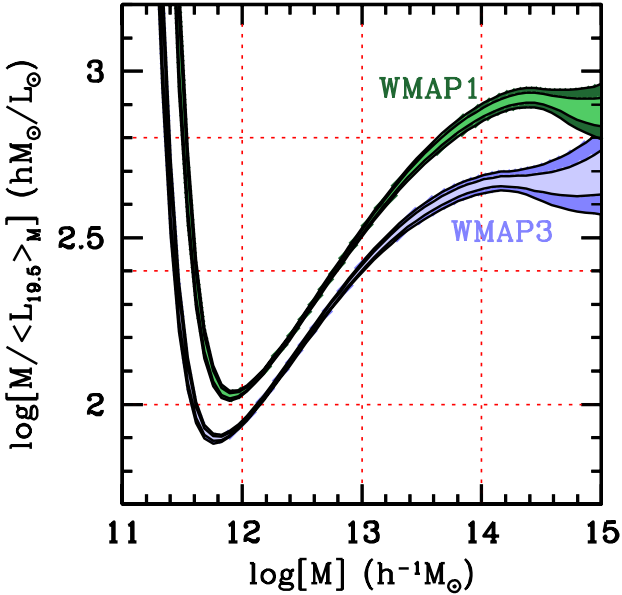,width=\hssize}}
\caption{The 68 and 95 percent confidence levels for the 
  mass-to-light ratios, $M/\langle  L_{19.5} \rangle_M$, obtained from
  the CLF MCMCs for the WMAP3 and WMAP1 cosmologies.}
\label{fig:MLrats}
\end{figure}

The fact that  both cosmologies allow an  (almost) equally good fit to
these  data,  despite the  relatively  large  differences in halo mass
function and halo bias, illustrates that  the abundance and clustering
properties of galaxies allow a fair  amount of freedom in cosmological
parameters.   However, as  demonstrated in van  den Bosch,  Mo \& Yang
(2003), the best-fit CLFs for  different cosmologies predict different
mass-to-light ratios as  function of halo mass.   This is evident from
Fig.~\ref{fig:MLrats}, which shows the mass-to-light ratios $M/\langle
L_{19.5} \rangle_M$   as function of halo  mass  inferred from our CLF
MCMCs  for the WMAP1  and WMAP3  cosmologies.   Here $\langle L_{19.5}
\rangle_M$ is  the average,  total   luminosity of all  galaxies  with
$^{0.1}M_r - 5 \log h  \leq -19.5$ that reside  in a halo of mass $M$,
which follows from the CLF according to
\begin{equation}
\langle L_{19.5} \rangle_M = 
\int_{L_{\rm min}}^{\infty} \Phi(L|M) \, L \, {\rm d} L\, ,
\end{equation}
with $L_{\rm  min}$  the  luminosity   corresponding to a    magnitude
$^{0.1}M_r  -  5 \log h =  -19.5$.   Clearly, the mass-to-light ratios
inferred for the WMAP1 cosmology are significantly higher than for the
WMAP3  cosmology (see also  van den Bosch  \etal 2007, where a similar
result was obtained using data from  the 2dFGRS). Hence, the abundance
and   clustering   properties of galaxies  can    be used to constrain
cosmological parameters, as long as one has independent constraints on
the mass-to-light ratios of dark matter  haloes.  This is exactly what
is provided by g-g lensing.  In the next section, we therefore use the
CLF models presented here to predict the g-g  lensing signal, which we
compare to SDSS data.

\section{Galaxy-Galaxy Lensing}
\label{sec:predGG}

\subsection{Model Predictions}
\label{sec:modpred}

In order  to compute the  ESD,  $\Delta\Sigma$, as a function  of  the
comoving  separation on the sky, $R$,  we proceed as follows. We start
by  calculating  the four  different terms  of the  galaxy-dark matter
cross    power  spectrum   defined     in  \S~\ref{sec:onehterm}   and
\S~\ref{sec:twohterm}.   Next we inverse   Fourier transform each  of
these terms using
\begin{equation}\label{4FT_back}
\xi^{\mu,{\rm x}}_{\rm g,dm}(r) = \frac{1}{2 \pi^2} \int  f_{\rm x}\, 
P^{\mu,{\rm x}}_{\rm g,dm}(k)\,  \frac{\sin(kr)}{kr}\, k^{2} \, {\rm d}k \ ,
\end{equation}
where `$\mu$' stands   for 1h or  2h, and  `x'  refers  to either  `c'
(centrals) or `s' (satellites). These are used to compute the
corresponding four terms of the surface density, $\Sigma^{\mu,{\rm x}}(R)$,
\begin{equation}\label{eq:Sigmaterms}
\Sigma^{\mu,{\rm x}}(R) = 2\overline{\rho}\int_{R}^{\infty} 
\xi^{\mu,{\rm x}}_{\rm g,dm}(r) \, 
{ r \rmd r \over \sqrt{r^2 - R^2}}\, .
\end{equation}
Note that we  are allowed to neglect the  contribution coming from the
constant      background     density,      $\bar{\rho}$,      (cf.     
equation~[\ref{eq:Sigma_approx}]) because  it will cancel  in defining
the  ESD  (this  is  known  in gravitational  lensing  theory  as  the
mass-sheet degeneracy).  The final ESD then simply follows from
\begin{eqnarray}\label{eq:4ESD}
\Delta \Sigma(R) & = & \Delta \Sigma^{\rm 1h, c}(R)+ \Delta \Sigma^{\rm 1h, s}(R) \nonumber \\
&+&\Delta \Sigma^{\rm 2h, c}(R)+\Delta \Sigma^{\rm 2h, s}(R) \, .
\end{eqnarray}
in which the relative weight of  each term is already included via the
central   and  satellite   fractions   in  the   definitions  of   the
corresponding power  spectra\footnote{Note that  our notation differs
slightly from that in Mandelbaum \etal (2006)}.
\begin{figure*}
\centerline{\psfig{figure=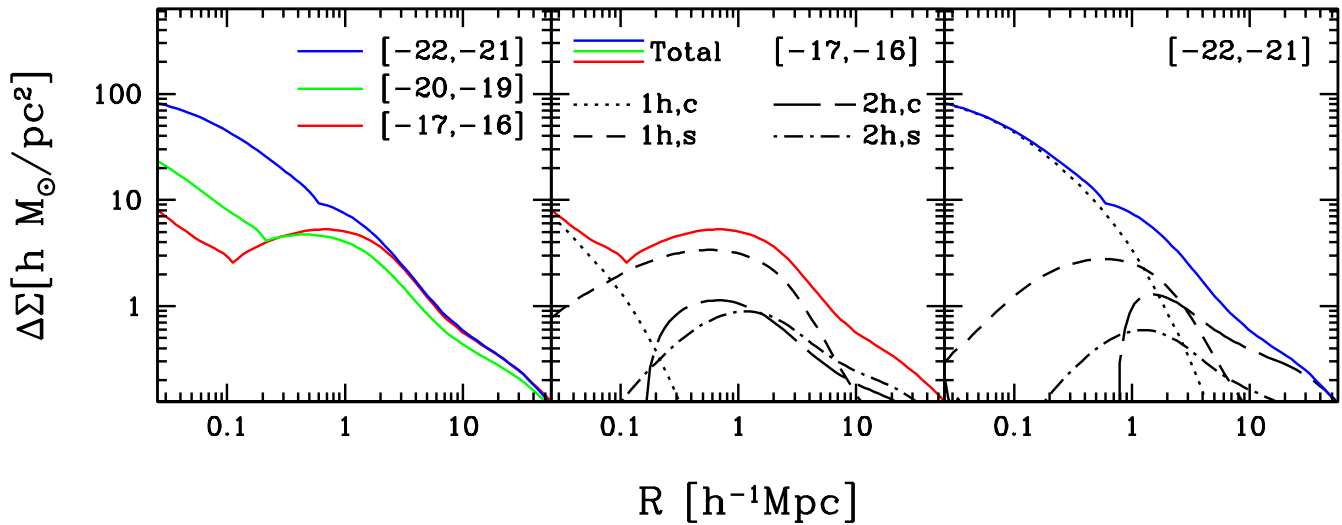,width=\hdsize}}
\caption{ The predicted ESD up to large scales
  ($R\sim~30 h^{-1}$Mpc) for three luminosity bins, as indicated.  The
  solid lines refer to the  total signal as predicted according to our
  model. The  dotted lines refer  to the 1-halo central  term, whereas
  the dashed lines refer to  the 1-halo satellite term. Note that they
  dominate the signal on different scales (see text).  The long dashed
  lines  refer  to  the  2-halo  central term.  It  rises  steeply  at
  relatively  large scales due  to our  halo exclusion  treatment (see
  Appendix).  The   2-halo  satellite  term  is   indicated  with  the
  dotted-dashed line.}
\label{fig:figLIV}
\end{figure*}

\begin{figure*}
\centerline{\psfig{figure=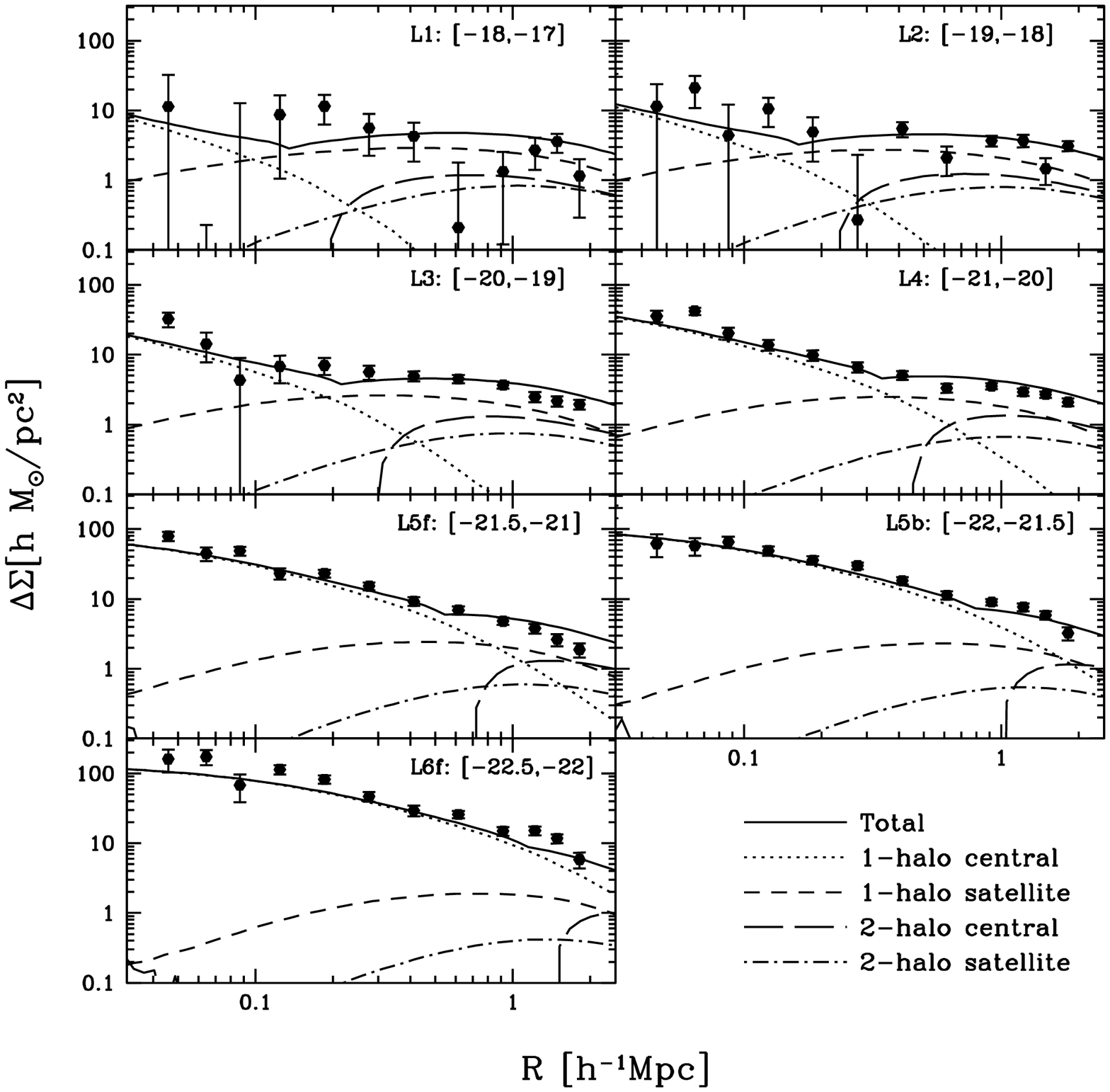,width=\hsize}}
\caption{The excess surface density  $\Delta \Sigma$ as a function of
  the comoving transverse separation $R$ is plotted for different bins
  in  luminosity of  the lens  galaxy (see  Table~4).  The  solid line
  represents the total  signal as predicted by the  model, data points
  and  error  bars  come  from  Seljak \etal  (2005),  see  text.  The
  different contributions to the  signal are also plotted.  The dotted
  line represents the 1-halo central term which obviously dominates at
  the smallest scales in all  cases.  Note that this term dominates on
  larger  and larger  scales  when brighter  galaxies are  considered,
  reflecting the idea  that brighter galaxies live on  average in more
  massive  haloes.  The  dashed line  represents the  1-halo satellite
  term  which  is  dominant  only  for  faint  galaxies  and  only  on
  intermediate  scales  (0.1-1 $h^{-1}$Mpc).   The  2-halo central  is
  plotted with  a long  dashed line and  it becomes relevant  on large
  scales ($R>1  h^{-1}$Mpc). Note that  the strong truncation  of this
  term at small scales is  due to our implementation of halo exclusion
  (see Appendix). The 2-halo satellite term (dotted-dashed line) never
  dominates but it can contribute up to 20\% of the total signal.}
\label{fig:all_contr}
\end{figure*}

Before comparing  the g-g lensing  predictions from our CLF  models to
actual  data,  we  first  demonstrate  how the  four  different  terms
contribute   to    the   total   ESD.    The    left-hand   panel   of
Fig.~\ref{fig:figLIV}  shows the  $\Delta\Sigma(R)$ obtained  from our
best-fit  CLF  model  for  the  WMAP3 cosmology  for  three  different
luminosity bins,  as indicated\footnote{Here, for  simplicity, we have
  used  the halo  mass function  and  halo bias  function computed  at
  $z=0$.  In  \S\ref{sec:lensingdata}, when  we compare our  models to
  data, we will  use the halo mass function and  halo bias function at
  the average redshift of the  lenses instead.}. Note that the fainter
luminosity  bins reveal  a  more `structured'  excess surface  density
profile, with a  pronouced `bump' at $R \sim 1  h^{-1} \Mpc$, which is
absent in  the $\Delta\Sigma(R)$ of the brighter  galaxies. The reason
for  this  is  evident  from  the  middle  and  right-hand  panels  of
Fig.~\ref{fig:figLIV},    which     show    the    contributions    to
$\Delta\Sigma(R)$ from  the four different  terms for the  faint ($-16
\geq  {^{0.1}M}_r  -  5\log  h   \geq  -17$)  and  bright  ($-21  \geq
{^{0.1}M}_r -  5\log h \geq  -22$) luminosity bins,  respectively.  In
both cases, the 1-halo central term dominates on small scales.  In the
case of the  faint galaxies, the 1-halo satellite  term dominates over
the radial range  $0.1 h^{-1} \Mpc \lta R \lta 5  h^{-1} \Mpc$, and is
responsible for  the pronounced bump  on intermediate scales.   In the
case of the bright bin, however, the 1-halo central term dominates all
the way  out to $R \sim  2 h^{-1} \Mpc$.   This owes to the  fact that
bright centrals  reside in more  massive haloes, which are  larger and
cause  a  stronger  lensing signal,  and  due  to  the fact  that  the
satellite fraction of brighter galaxies is smaller.  The fact that the
1-halo  satellite term peaks  at intermediate  scales, rather  than at
$R=0$, owes  to the fact that $\Delta\Sigma^{\rm  1h,s}(R)$ reflects a
convolution  of the  host halo  mass density  profile with  the number
density distribution of satellite  galaxies.  On large scales ($R \gta
3 h^{-1} \Mpc$), which roughly reflects two times the virial radius of
the  most massive  dark matter  haloes, the  ESD is  dominated  by the
2-halo  terms.    Note  that  the  faint  galaxies,   with  $-16  \geq
{^{0.1}M}_r - 5\log h \geq -17$,  have the same large scale ESD as the
bright  galaxies with  $-21  \geq  {^{0.1}M}_r -  5\log  h \geq  -22$,
indicating that they have similar values for their bias.  This owes to
the fact that  many of the faint galaxies  are satellites which reside
in massive  haloes. Note also that  the 2-halo central  term reveals a
fairly abrupt truncation at  small radii, which owes to halo-exclusion
(see Appendix).  This truncation also  leaves a signature in the total
lensing signal,  which is more  pronounced for the fainter  lenses. We
caution,  however,  that  the  {\it  sharpness}  of  this  feature  is
partially  an  artefact  due  to  our  approximate  implementation  of
halo-exclusion. Nevertheless,  it is clear  from Fig.~\ref{fig:figLIV}
that  the   excess  surface   densities  obtained  from   g-g  lensing
measurements contain a wealth of information regarding the galaxy-dark
matter connection.

\subsection{Data}
\label{sec:lensingdata}

The g-g lensing data used here is described in Seljak \etal (2005) and
Mandelbaum \etal (2006) and has been kindly  provided to us by
R.   Mandelbaum.   The  catalogue  of  lenses  consists  of  $351,507$
galaxies with  magnitudes $-17 \geq {^{0.1}M}_r  - 5 \log  h \geq -23$
and redshifts $0.02  < z < 0.35$ taken from  the main galaxy catalogue
of the SDSS Data Release 4 (Adelman-McCarthy \etal 2006).  This sample
is split  in 7 luminosity  bins (see Table~4),  and for each  of these
luminosity    bins    the    excess    surface    density    profiles,
$\Delta\Sigma(R)$,  have  been  determined  from measurements  of  the
shapes of more than 30 million  galaxies in the SDSS imaging data down
to an apparent $r$-band magnitude of $r=21.8$.  The resulting data are
shown as  solid dots  with errorbars in  Fig.~\ref{fig:all_contr}.  We
refer  the reader  to Mandelbaum  \etal (2005b,  2006) for  a detailed
description of the  data and of the methods used  to determine the ESD
profiles.
\begin{table}
\label{tabSDSS}
\caption{Luminosity bins of the SDSS g-g lensing data}
\begin{tabular}{lcc}
\hline\hline
  ID & $^{0.1}M_r - 5 \log h$ & $\langle z \rangle$ \\
 (1) & (2)                    & (3)                 \\
\hline
 L1  & $(-18.0,-17.0]$ & 0.032 \\
 L2  & $(-19.0,-18.0]$ & 0.047 \\
 L3  & $(-20.0,-19.0]$ & 0.071 \\
 L4  & $(-21.0,-20.0]$ & 0.10 \\
 L5f & $(-21.5,-21.0]$ & 0.14 \\
 L5b & $(-22.0,-21.5]$ & 0.17 \\
 L6f & $(-22.5,-22.0]$ & 0.20 \\
\hline\hline
\end{tabular}
\begin{minipage}{\hsize}
  Luminosity  bins of  the   lenses. Column  (1) lists  the  ID of each
  luminosity bin, following the  notation of Mandelbaum \etal  (2006).
  Column (2) indicates the magnitude range of each luminosity bin (all
  magnitudes are K$+$E corrected to $z=0.1$). Column (3) indicates the
  mean redshift of the lenses  in each luminosity bin. See  Mandelbaum
  \etal (2006) for details.
\end{minipage}
\end{table}

\subsection{Results for the WMAP3 Cosmology}
\label{sec:fiducial}

Using   the   methodology   outlined  in   \S\ref{sec:gglensing}   and
\S\ref{sec:modpred},  we  now use  the  CLF  for  the WMAP3  cosmology
obtained in \S\ref{sec:CLF} to predict  the g-g lensing signal for the
7  luminosity bins  listed in  Table~4.   For each  luminosity bin  we
compute the  ESD profile, $\Delta\Sigma(R)$,  at the mean  redshift of
the sample, i.e., we use the halo mass function, $n(M)$, the halo bias
function, $b(M)$, and the  non-linear power spectrum, $P_{\rm dm}^{\rm
  NL}(k)$ that  correspond to  the mean redshift  listed in  the third
column  of   Table~4.   We  have  verified,   though,  that  computing
$\Delta\Sigma(R)$ simply  at $z=0$ instead has a  negligible impact on
the results.

The results  are  shown in  Fig.~\ref{fig:all_contr},  where the solid
dots with  errorbars correspond to  the SDSS data  and the solid lines
are the predictions of  our best-fit CLF  model (whose  parameters are
listed in  Table~3).  Note that  this model fits the data remarkably
well, which is  quantified  by the  fact that the  reduced $\chi^2$ is
$3.1$.  We emphasize  that there are  no free parameters here: the ESD
has been computed using the CLF that has been constrained using the LF
and the clustering data.  The good agreement between model and lensing
data  thus  provides independent  support    for the halo   occupation
statistics described  by our  WMAP3  CLF model, in particular  for the
mass-to-light ratios and satellite fractions,  which have an important
impact on the lensing signal.

The different curves in each of the panels in Fig.~\ref{fig:all_contr}
show the contribution  to the lensing signal due  to the four separate
terms: $\Delta \Sigma^{\rm  1h,c}$ (dotted lines), $\Delta \Sigma^{\rm
  1h,s}$ (short-dashed lines), $\Delta \Sigma^{\rm 2h,c}$ (long-dashed
lines),  and  $\Delta  \Sigma  ^{\rm 2h,s}$  (dot-dashed  lines).   In
agreement with the examples shown in Fig.~\ref{fig:figLIV}, the 1-halo
central  term becomes  increasingly  more dominant  for more  luminous
lenses.  In fact,  in the brightest luminosity bin  (L6f) it dominates
over the entire radial range  probed. In the low-luminosity bins, most
of  the  observed  lensing signal  at  $R  \gta  200 h^{-1}  \kpc$  is
dominated  by the  1-halo  satellite  term. The  fact  that our  model
accurately  fits  the  data,  thus supports  the  satellite  fractions
inferred from our  CLF model, and shown in  the upper right-hand panel
of Fig.~\ref{fig:CLFres3}.

Both Seljak \etal  (2005) and Mandelbaum \etal (2006)  did not account
for the  contributions of  the 2-halo terms  in their analyses  of the
galaxy-galaxy lensing signal.  Our  model indicates that, although the
2-halo terms never  dominate the total signal, they  can contribute as
much as 50 percent at large  radii ($R \simeq 1 h^{-1} \Mpc$). We thus
conclude that the 2-halo terms cannot simply be ignored.

\subsection{Comparison with the WMAP1 Cosmology}
\label{sec:wmap1res}

\begin{figure*}
\centerline{\psfig{figure=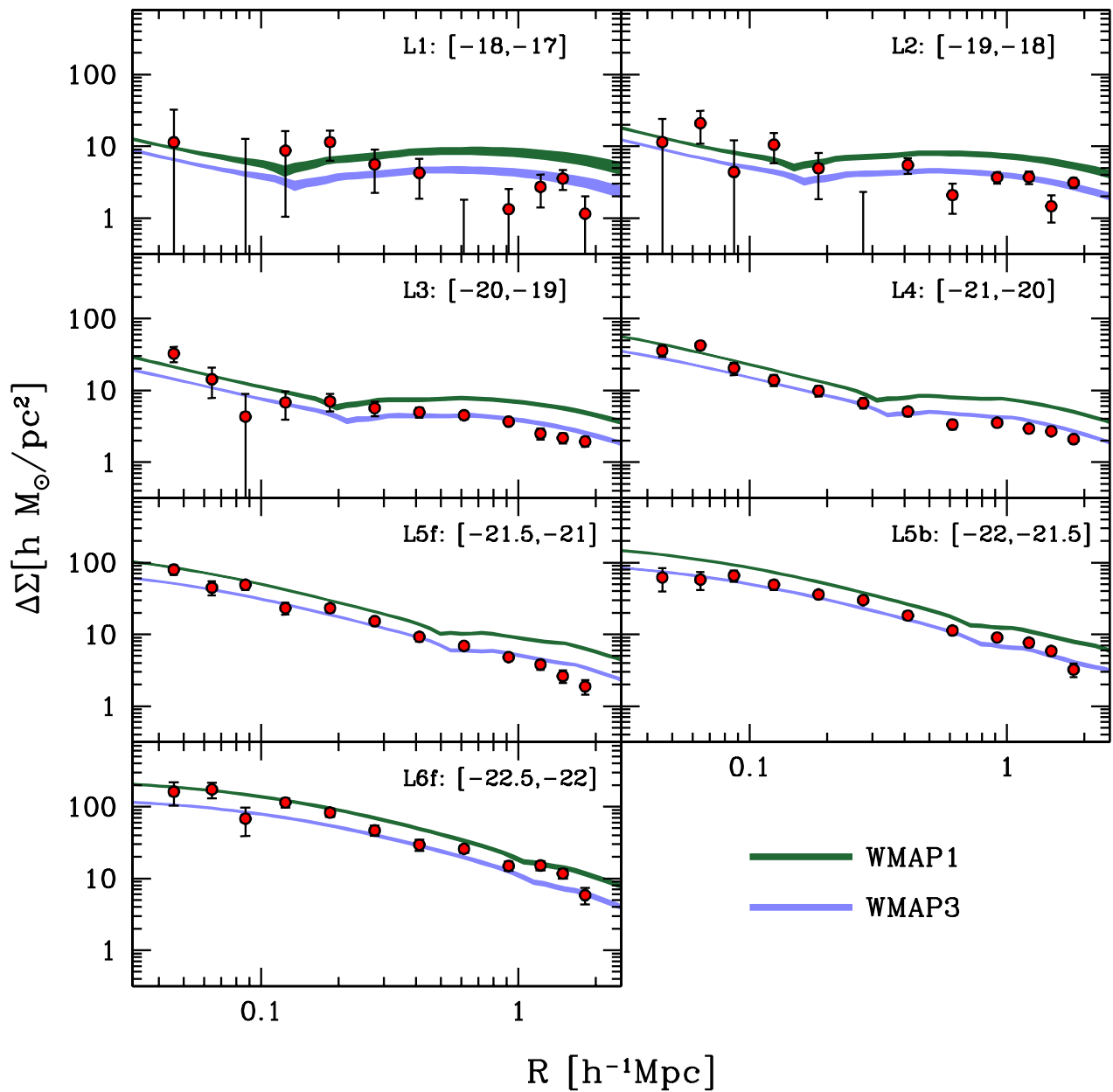,width=\hsize}}
\caption{
  The  predictions for the  lensing signal,  $\Delta \Sigma  (R)$, are
  shown for  two different sets of cosmological  parameters (WMAP1 and
  WMAP3, see  text). The green  (blue) shaded area corresponds  to the
  95\%  confidence  level  of  the  WMAP1 (WMAP3)  model.  Note  that,
  although the  cosmological parameters of these  two cosmologies only
  differ  by $\lta  20$  percent (see  Table~1),  the ESD  predictions
  are very different, and can easily be discriminated.}
\label{fig:cosmocomp}
\end{figure*}

As shown  in \S\ref{sec:CLFres}, the WMAP3 and  WMAP1 cosmologies both
allow  a good  fit to  the  clustering data,  luminosity function  and
galaxy  group   results.  However,  the   corresponding  CLFs  predict
mass-to-light  ratios  that are  significantly  different.  Since  the
galaxy-galaxy lensing signal is  very sensitive to these mass-to-light
ratios,  it is  to be  expected  that our  WMAP3 and  WMAP1 CLFs  will
predict  significantly different  ESD  profiles, thus  allowing us  to
discriminate between these two cosmologies.

Fig.~\ref{fig:cosmocomp}  shows the 95  percent confidence  levels for
$\Delta\Sigma(R)$  obtained from  our  CLF MCMCs  for  both the  WMAP3
(blue) and WMAP1 (green)  cosmologies. Indeed, as anticipated, for the
WMAP1  cosmology   we  obtain   excess  surface  densities   that  are
significantly higher than for the  WMAP3 cosmology, in accord with the
higher   mass-to-light   ratios   (cf.    Fig.~\ref{fig:MLrats}).    A
comparison with the SDSS data  clearly favors the WMAP3 cosmology over
the WMAP1 cosmology.   In fact, for the latter  our best-fit CLF model
yields a  reduced $\chi^2$ of $29.5$,  much larger than  for the WMAP3
cosmology  ($\chi^2_{\rm  red}=3.1$).    Note  that  the  cosmological
parameters for  these two  cosmologies are very  similar: $\Omega_{\rm
  m}$ and $\sigma_8$ differ only  by $\sim 20$ percent (in addition to
a $\sim 5$ percent difference in $n$).  Yet, we can very significantly
favor  one  cosmology  over   the  other.   This  indicates  that  the
combination of clustering data and g-g lensing data can be used to put
tight  constraints on  cosmological parameters.   A  detailed analysis
along these lines is deferred  to a forthcoming paper (Cacciato \etal,
in preparation).
\begin{figure}
\centerline{\psfig{figure=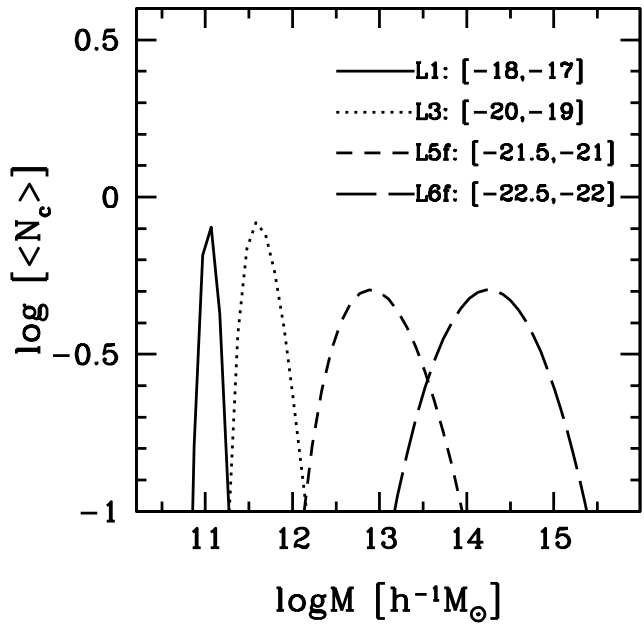,width=\hssize}}
\caption{The average number of central galaxies as a function of halo
  mass obtained from  our best-fit CLF for the  WMAP3 cosmology.  This
  is equivalent  to the probability $\calP_c(M \vert  L_1,L_2)$ that a
  central galaxy  with $L_1 \leq  L \leq L_2$  is hosted by a  halo of
  mass $M$.  Results are shown  for four different luminosity bins, as
  indicated.  Note that brighter  centrals reside, on average, in more
  massive haloes. In addition, the width of $\calP_c(M \vert L_1,L_2)$
  also increases with luminosity.}
\label{fig:HOS}
\end{figure}
\begin{figure*}
\centerline{\psfig{figure=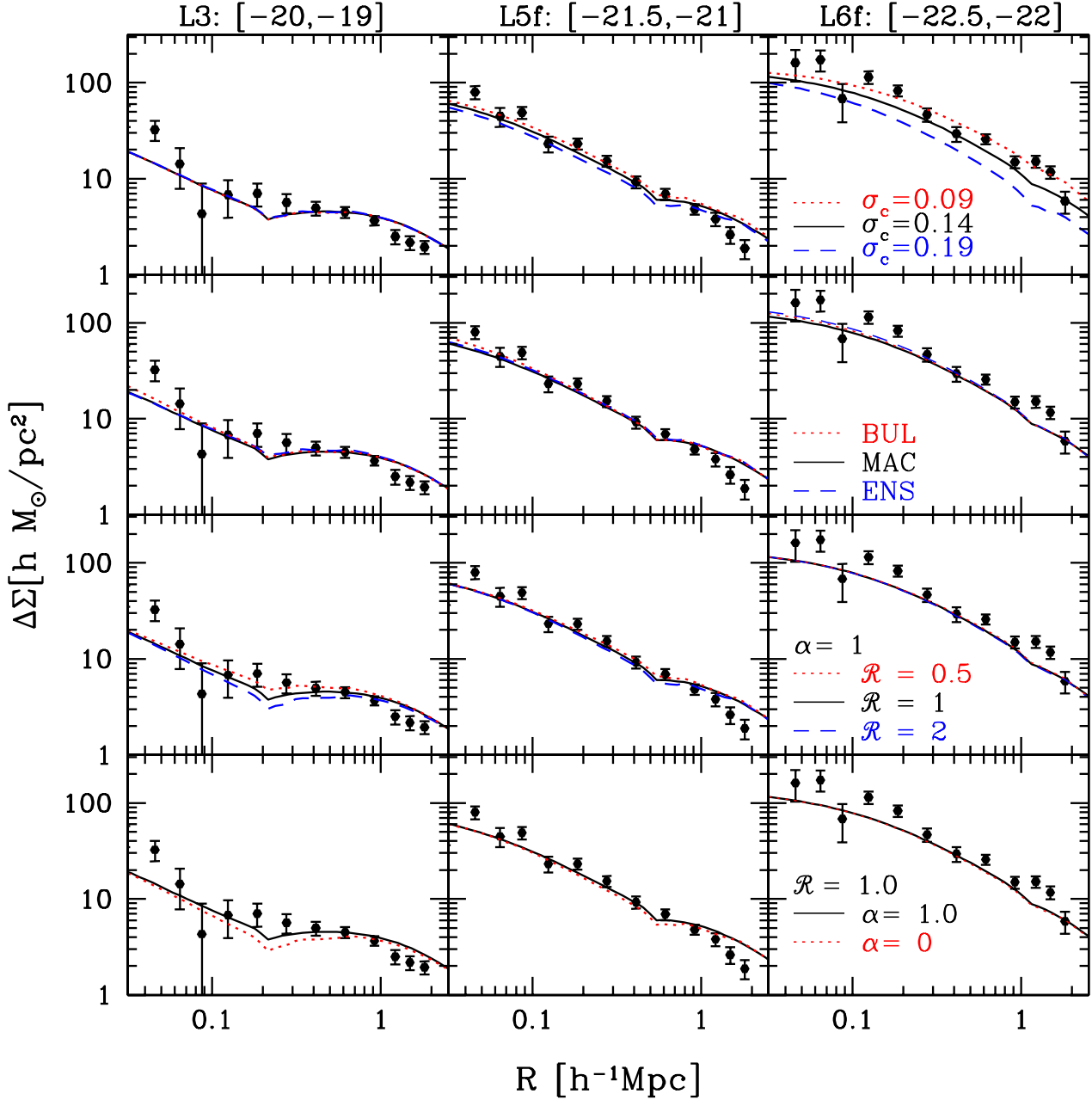,width=\hsize}}
\caption{The impact of various model parameters on $\Delta \Sigma(R)$.
  Results are shown for three luminosity bins, as indicated at the top
  of each  column.  In  each panel the  solid line corresponds  to our
  fiducial  model (the  best-fit  CLF model  for  the WMAP3  cosmology
  presented in Fig.~\ref{fig:all_contr}),  while the dotted and dashed
  lines  correspond  to models  in  which  we  have only  changed  one
  parameter or  model ingredient.  {\it  Upper panels:} the  impact of
  changes in  the parameter $\sigma_\rmc$, which  describes the amount
  of  scatter  in  $\Phi_\rmc(L|M)$ (see  equation~[\ref{eq:phi_c}]).  
  {\it Second  row from the  top:} the impact  of changes in  the halo
  concentration,  $c_{\rm dm}(M)$.   In particular,  we  compare three
  models  for the  mass  dependence of  $c_{\rm  dm}$: Macci\`o  \etal
  (2007;  MAC),  Bullock  \etal  (2001;  BUL),  and  Eke \etal (2001; ENS).  {\it Third  row from the top:} the impact of
  changes  in   $\calR  =  c_{\rm  dm}/c_\rms$,   which  controls  the
  concentration of the radial number density distribution of satellite
  galaxies relative to  that of the dark matter.   {\it Lower panels:}
  The impact of changes in $\alpha$, which specifies the central slope
  of  the radial number  density distribution  of satellite  galaxies. 
  See text for a detailed discussion.}
\label{fig:cimpact}
\end{figure*}

\section{Model Dependencies}
\label{sec:moddep}

Although our computation of the   g-g lensing signal does not  involve
any free  parameters (these are already  constrained by the clustering
data), a number  of assumptions are  made.  In particular, haloes  are
assumed to   be spherical and  to  follow a  NFW density distribution,
central galaxies are assumed to reside  exactly at the center of their
dark matter  haloes, and satellite   galaxies are assumed to  follow a
radial number density distribution that has the same shape as the dark
matter mass distribution. In  addition, we made assumptions  regarding
the functional form of the CLF. Although most of these simplifications
are reasonable,  and have support  from independent studies,  they may
have  a  non-negligible impact on  the predictions  of the g-g lensing
signal. If this is the  case, they will affect  the reliability of the
cosmological  constraints inferred from the   data. In this section we
therefore investigate how  strongly  our model  predictions  depend on
these oversimplified assumptions.

Some of these dependencies  were already investigated in our companion
paper (Li \etal 2008). In particular, we have shown that the fact that
realistic dark  matter haloes are ellipsoidal,  rather than spherical,
can  be  safely ignored  (i.e.,  its impact  on  the  ESD profiles  is
completely negligible). On the other hand, if central galaxies are not
located exactly  at the center of  their dark matter  haloes, this may
have a non-negligible impact on the lensing signal on small scales ($R
\lta 0.1 h^{-1} \Mpc$).  Fortunately, for realistic amplitudes of this
offset (van  den Bosch  \etal 2005b), the  effect is fairly  small and
only restricted to  the most luminous galaxies (see  Li \etal 2008 for
details).

Below we investigate three  additional model dependencies: the scatter
in  the relation  between light  and mass,  the concentration  of dark
matter haloes, and the radial number density distribution of satellite
galaxies. To that  extent we compare our fiducial  model (the best-fit
CLF model for the WMAP3 cosmology presented above), to models in which
we  change  only  one  parameter.  

\subsection{Scatter in the $L_{\rm c}-M$ relation}
\label{sec:impactScatter}

An important improvement  of our halo occupation  model over that used
by Seljak \etal  (2005) and Mandelbaum \etal (2006)  is that  we allow
for scatter in the relation between light and mass.  In particular, we
model the  probability function $\calP_c(L \vert   M) = \Phi_c(L \vert
M)$ as a log-normal with a scatter, $\sigma_c$, that  is assumed to be
independent of halo mass.  As demonstrated in More \etal (2008b), this
assumption is consistent  with  the kinematics of satellite  galaxies,
and it is supported  by semi-analytical  models for galaxy  formation.
Note, though, that this does not imply  that the scatter in $\calP_c(M
\vert L)$, which is the probability function  which actually enters in
the computation of the g-g lensing  signal, is also constant. In fact,
it is not.   This  is illustrated in Fig.~\ref{fig:HOS},   which shows
$\calP_c(M  \vert L_1,L_2)$ of our fiducial  model for four luminosity
bins.  Two  trends  are evident:  more   luminous centrals reside,  on
average, in more massive haloes,  and they  have  a larger scatter  in
halo masses.  As discussed in More, van  den Bosch \& Cacciato (2008),
the  fact  that the   scatter    in $\calP_\rmc(M|L)$ increases   with
luminosity simply  owes to  the  fact that   the slope of  $L_\rmc(M)$
becomes shallower with increasing $M$ (see the lower right-hand panels
of Figs.~\ref{fig:CLFres3} and~\ref{fig:CLFres1}).  As is evident from
Fig.~\ref{fig:HOS}, this  is  a  strong  effect,  with the  scatter in
$\calP_\rmc(M|L)$ becoming extremely large   at the bright end.   Note
that this scatter is not dominated by the width of the luminosity bin.
Hence, even if one were able to use infinitesimally narrow luminosity
bins when stacking lenses, the scatter in $\calP_\rmc(M|L)$ will still
be very appreciable.

As first  shown by  Tasitsiomi \etal (2004),  scatter in  the relation
between light and mass can have a very significant impact on the ESDs.
This is  demonstrated in  the upper panels  of Fig.~\ref{fig:cimpact},
which show  the impact  on $\Delta\Sigma(R)$ of  changing $\sigma_{\rm
  c}$   by   $0.05$   compared   to   our  best-fit   CLF   value   of
$\sigma_\rmc=0.14$;  all  other parameters  are  kept  fixed at  their
fiducial   values  (see   Table~3).   Note   that  these   changes  in
$\sigma_{\rm c}$ have a negligible impact on $\Delta\Sigma(R)$ for the
low  luminosity bins.  At  the bright  end, however,  relatively small
changes   in  $\sigma_\rmc$   have  a   very  significant   impact  on
$\Delta\Sigma(R)$.   In particular, increasing  the amount  of scatter
{\it reduces}  the ESD. This  behavior owes to  the shape of  the halo
mass function. Increasing the scatter adds both low mass and high mass
haloes to the distribution, and the overall change in the average halo
mass  depends  on the  slope  of  the  halo mass  function.   Brighter
galaxies live  on average in more  massive haloes where  the halo mass
function is  steeper.  In  particular, when the  average halo  mass is
located at the exponential tail of the halo mass function, an increase
in the  scatter adds  many more low  mass haloes than  massive haloes,
causing a drastic shift in the average halo mass towards lower values.
On the other hand, fainter galaxies live in less massive haloes, where
the slope of the halo  mass function is much shallower.  Consequently,
a change  in the scatter  does not cause  a significant change  in the
average mass.

Clearly, if the  g-g lensing signal is used  to constrain cosmological
parameters,  it is  important  that one  has  accurate constraints  on
$\sigma_\rmc$.    From   the    clustering   analysis   presented   in
\S\ref{sec:CLFres},  we obtain  $0.14 \pm  0.01$ (for  both  WMAP1 and
WMAP3).   This is  in  good agreement  with  previous studies:  Cooray
(2006),  using  a  CLF  to   model  the  SDSS  $r$-band  LF,  obtained
$\sigma_\rmc  =  0.17^{+0.02}_{-0.01}$.  YMB08,  using  a SDSS  galaxy
group  catalogue, obtained  $\sigma_\rmc =  0.13 \pm  0.03$,  and More
\etal (2008b), using the kinematics  of satellite galaxies in the SDSS,
find $\sigma_\rmc  = 0.16  \pm 0.04$ (all  errors are  68\% confidence
levels).  Although it is reassuring that very different methods obtain
values that are in such good agreement, it is clear that the remaining
uncertainty  may  have a  weak  impact  on  our ability  to  constrain
cosmological  parameters.  Fortunately, the  scatter only  impacts the
results at the bright end, so that one can always check the results by
removing data from the brightest luminosity bins.

\subsection{The dark matter halo concentration}

The g-g lensing  signal on  small scales  reflects the  projected mass
distribution of the   haloes hosting the lensing  galaxies. Therefore,
the detailed  shape of $\Delta\Sigma(R)$ on  small scales is sensitive
to the mass distribution of dark matter haloes.  In our model, we have
assumed that dark  matter   haloes  follow NFW profiles,   which   are
characterized by their  concentration parameters,  $c_{\rm dm}$.  Halo
concentrations  are known to depend on  both halo  mass and cosmology,
and various analytical models have  been  developed to describe  these
dependencies (Navarro \etal 1997;  Bullock \etal 2001; Eke, Navarro \& Steinmetz 2001;
Neto \etal 2007;   Macci\`o \etal 2007, 2008).    Unfortunately, these
models make slightly different  predictions for the mass dependence of
$c_{\rm dm}$  (mainly due to  the fact that  the numerical simulations
used to calibrate the  models covered different limited mass  ranges).
In Li \etal  (2008), we  have  shown that changing  $c_{\rm  dm}$ by a
factor  of two has  a very large  impact on the ESD profiles. However,
this   is  much larger  than   the  typical discrepancies between  the
different models for $c_{\rm  dm}(M)$.   The second  row of panels  in
Fig.~\ref{fig:cimpact} shows  $\Delta\Sigma(R)$ obtained for  three of
these   models: the  solid  lines  (labelled MAC)  corresponds  to our
fiducial model for which we have used  the $c_{\rm dm}(M)$ relation of
Macci\`o \etal  (2007).   The dotted lines   (labelled BUL) and dashed
lines (labelled  ENS) correspond to the  $c_{\rm  dm}(M)$ relations of
Bullock \etal  (2001) and  Eke  \etal (2001), respectively.    The BUL
model predicts  halo concentrations that are  about  15 percent higher
than for the MAC model.  The ENS model predicts a $c_{\rm dm}(M)$ that
is somewhat shallower than the BUL and MAC models.  As is evident from
Fig.~\ref{fig:cimpact},  though,  the  results  based  on these  three
different models are very similar.   We thus conclude that our results
are robust to uncertainties in the relation between halo mass and halo
concentration.

\subsection{Number density of satellite galaxies}
\label{sec:impactSat}

In our modelling of the  g-g lensing signal,  we have assumed that the
number density distribution of satellite galaxies  can be described by
a generalised  NFW profile  (eq.~[\ref{eq:generalisedNFW}]),  which is
parameterized by two free  parameters:  $\alpha$ and $\calR$.   In the
models presented above, we have assumed that $\alpha=\calR=1$, so that
the number density distribution  of satellite galaxies has exactly the
same  shape as  the  dark   matter  distribution.    As discussed   in
\S\ref{sec:ingredients}, though, there is observational evidence which
suggests   that satellite galaxies   are  spatially   anti-biased with
respect to the  dark matter (i.e., their  radial distribution is  less
concentrated than that of the dark matter).  This is also supported by
numerical simulations, which show   that dark matter subhaloes  (which
are believed   to    host satellite  galaxies)   are   also  spatially
anti-biased with respect to the  dark matter (e.g., Moore \etal  1999,
De Lucia 2004).

The panels in the third  row of Fig.~\ref{fig:cimpact} show the impact
of   changing  the   concentration  of   the  radial   number  density
distribution of satellite galaxies.  In particular, we compare the ESD
profiles obtained  for our  fiducial model ($\calR=1.0$,  solid lines)
with  models  in  which  $\calR=0.5$ (dotted  lines)  and  $\calR=2.0$
(dashed  lines). Recall  that  $\calR =  c_{\rm  dm}/c_\rms$, so  that
$\calR > 1$ ($\calR < 1$) corresponds to satellite galaxies being less
(more) centrally concentrated than  the dark matter. Note that changes
in  $\calR$ have  a  negligible effect  on  $\Delta\Sigma(R)$ for  the
bright luminosity bins.  This simply owes  to the fact that the ESD of
bright  lenses is  completely  dominated by  the  1-halo central  term
(i.e., the satellite  fraction of bright galaxies is  very small). For
the  fainter  luminosity  bins,  however, an  increase  (decrease)  in
$\calR$   causes  a  decrease   (increase)  in   $\Delta\Sigma(R)$  on
intermediate scales  ($0.1 h^{-1}  \Mpc \lta R  \lta 1  h^{-1} \Mpc$),
which is the scale on  which the 1-halo satellite term dominates.  The
effect, though, is fairly  small (typically smaller than the errorbars
on the data points).
 
The last  row of Fig.~\ref{fig:cimpact}  shows the impact  of changing
the central  slope, $\alpha$, of  $n_\rms(r)$.  If the  number density
distribution of  satellite galaxies  has a central  core ($\alpha=0$),
rather than a  NFW-like cusp ($\alpha=1$), it has  a similar impact on
the  lensing   signal  as  assuming  a   less  centrally  concentrated
$n_\rms(r)$.  In fact, the ESD profiles for $(\alpha,\calR)=(0.0,1.0)$
are very  similar to  those for $(\alpha,\calR)=(1.0,2.0)$.   The main
conclusion, though, is that our  results are not very sensitive to the
exact form  of $n_\rms(r)$  (see also Yoo  \etal 2006).   Clearly, our
conclusion  that the WMAP3  cosmology is  strongly preferred  over the
WMAP1  cosmology  is  not  affected  by uncertainties  in  the  radial
distribution of satellite galaxies.

\section{Conclusions}
\label{sec:conclusions}

Galaxy  clustering  and galaxy-galaxy  lensing  probe the  galaxy-dark
matter connection in complementary ways.  Since the clustering of dark
matter  haloes depends  on cosmology,  the halo  occupation statistics
inferred  from  the observed  clustering  properties  of galaxies  are
degenerate  with   the  adopted  cosmology.   Consequently,  different
cosmologies  imply  different  mass-to-light  ratios for  dark  matter
haloes.  Galaxy-galaxy lensing, on the other hand, yields {\it direct}
constraints on the actual mass-to-light  ratios of dark matter haloes. 
Combined, clustering  and lensing  therefore offer the  opportunity to
constrain cosmological parameters.

Although the  advent of  wide and deep  surveys has resulted  in clear
detections of galaxy-galaxy lensing,  a proper interpretation of these
data in terms of the link  between galaxies and dark matter haloes has
been hampered by the fact that the lensing signal can only be detected
when stacking the  signal of many lenses. Since  not all lenses reside
in  haloes of the  same mass,  the resulting  signal is  a non-trivial
average of the  lensing signal due to haloes  of different masses.  In
addition, central  galaxies (those  residing at the  center of  a dark
matter halo)  and satellite galaxies (those orbiting  around a central
galaxy)  contribute very  different  lensing signals,  even when  they
reside in haloes of the same mass (e.g., Yang \etal 2006). This has to
be  properly  accounted  for,  and  requires  knowledge  of  both  the
satellite fractions and of  the spatial number density distribution of
satellite galaxies within their dark matter haloes.

In this  paper,   we  model  galaxy-galaxy   lensing  with   the  CLF,
$\Phi(L|M)$,    which describes the   average   number of galaxies  of
luminosity $L$ that reside in a halo of mass  $M$. This CLF is ideally
suited   to  model  galaxy-galaxy  lensing.    In   particular, it  is
straightforward to  account for the fact  that there is scatter in the
relation between  the luminosity of a  central galaxy and  the mass of
its dark  matter halo. This  represents an improvement with respect to
previous attempts  to model the  g-g lensing signal obtained  from the
SDSS, which  typically ignored this  scatter (e.g.  Seljak \etal 2005;
Mandelbaum \etal 2006).  However,  in agreement with Tasitsiomi  \etal
(2004), we have demonstrated that the scatter  in this relation has an
important impact on the  g-g lensing signal  and cannot be ignored. We
also improved upon previous studies  by modelling the 2-halo term (the
contribution to  the lensing signal   due   to the mass   distribution
outside of the halo hosting the lens galaxy), including an approximate
treatment for halo exclusion.

Following  Cooray \& Milosavljevi\'c (2005),  we split  the CLF in two
components: one for  the central galaxies  and one for the satellites.
This facilitates a proper treatment  of their respective contributions
to the g-g  lensing  signal.  The  functional  forms for the two   CLF
components are motivated by results obtained by Yang \etal (2008) from
a large  galaxy group  catalogue.  For  a  given  cosmology, the  free
parameters  of the CLF  are constrained using the luminosity function,
the correlation lengths as function of luminosity, and some properties
extracted from the group  catalogue.  We  have performed our  analysis
for two different $\Lambda$CDM cosmologies: the WMAP1 cosmology, which
has  $\Omega_{\rm m}=0.3$ and  $\sigma_8=0.9$  and the WMAP3 cosmology
with    $\Omega_{\rm m}=0.238$   and    $\sigma_8=0.744$.    For  both
cosmologies we   have   obtained CLFs that    can   accurately fit the
abundances and clustering  properties of SDSS galaxies. However, these
CLFs   predict mass-to-light ratios    that are very  different.  This
reflects  the  degeneracy  between   cosmology   and halo   occupation
statistics alluded to above. In order to break this degeneracy, we use
these CLFs to predict the g-g  lensing signal (with no additional free
parameters), which  is compared to   the SDSS data  obtained by Seljak
\etal (2005) and Mandelbaum \etal (2006).  While the WMAP3 CLF
predictions are in excellent agreement with the data,  the CLF for the
WMAP1 cosmology predicts excess surface densities that are much higher
than observed.  Although the  cosmological parameters of the WMAP1 and
WMAP3 cosmologies only differ at the 20 percent level, the combination
of   clustering  and lensing allows   us  to strongly  favor the WMAP3
cosmology over the WMAP1 cosmology.  In a  companion paper by Li \etal
(2008), we use a completely  different technique to model g-g lensing,
but nevertheless reach exactly the same conclusion.

In order  to test the  robustness of our  results we have  performed a
number of tests. In particular, we have shown that small uncertainties
in the expected concentrations of dark matter haloes, or in the radial
number density  distributions of satellite galaxies, only  have a very
small impact  on the predicted  lensing signal. In  addition, although
our  treatment  of  halo   exclusion  is  only  approximate,  we  have
demonstrated that it is sufficiently accurate. Finally, as shown by Li
\etal (2008),  making the  oversimplified assumption that  dark matter
haloes  are spherical rather  than ellipsoidal  also has  a negligible
impact on the  lensing predictions.  We thus conclude  that our method
yields accurate and reliable predictions for g-g lensing.

To  summarize,   as  already  discussed  by  Yoo   \etal  (2006),  the
combination  of  clustering and  lensing  can  be  used to  put  tight
constraints on  cosmological parameters. In  this pilot study  we have
shown  that current  data  from  the SDSS  strongly  favors the  WMAP3
cosmology  over the  WMAP1 cosmology.   In a  follow-up paper  we will
present a more detailed  analysis of the cosmological constraints that
can be obtained using this technique.


\section*{Acknowledgements}

M.C.  acknowledges  R.  Mandelbaum for  providing the lensing  data in
electronic  format and  for  her kind  cooperation.  M.C. also  thanks
Alexie Leauthaud for useful discussions, as well as Nikhil Padmanabhan
and  Joanne  Cohn   for  useful  comments  during  the   visit  at  UC
Berkeley/LBNL.



\appendix
\section{Halo exclusion}

\begin{figure}
  \centerline{\psfig{figure=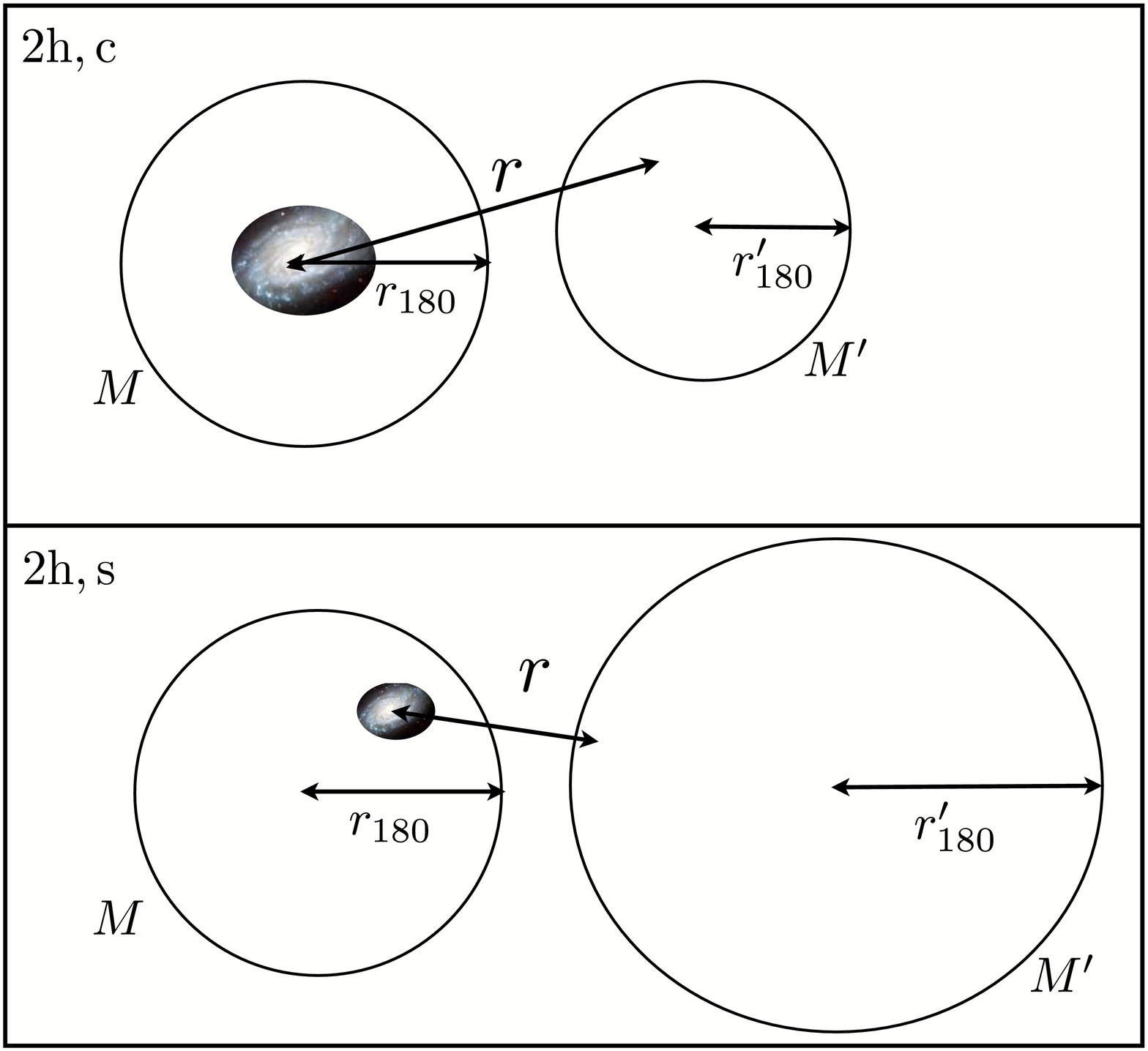,width=0.95\hssize}}
\caption{Illustration of halo exclusion. The upper panel shows two
  haloes, of  masses $M$ and  $M'$, and corresponding  radii $r_{180}$
  and $r'_{180}$, respectively.  The halo  of mass $M$ hosts a central
  galaxy. Since  two haloes cannot  overlap, this central  galaxy does
  not  contribute  any  signal  to  the 2-halo  central  term  of  the
  galaxy-dark matter cross correlation function on scales $r<r_{180}$.
  In the case  of the 2-halo satellite term,  illustrated in the lower
  panel, there is  still a contribution even on  very small scales ($r
  \ll r_{180}$), simply because satellite galaxies can reside near the
  edge of the halo.}
\label{fig:sketchHE}
\end{figure}

By  definition,  the 2-halo  terms  of  the  galaxy-dark matter  cross
correlation, $\xi_{\rm g,dm}(r)$, only considers pairs of galaxies and
dark  matter particles  that  reside in  different  haloes. Since  two
haloes can not  overlap spatially, this implies that  the 2-halo terms
given  by  Eqs.\,(\ref{eq:fcP2hc}) and  (\ref{eq:fsP2hs})  need to  be
modified  to take  account of  halo  exclusion.  The  concept of  halo
exclusion  is illustrated  in Fig.~\ref{fig:sketchHE}  for  the 2-halo
central and  2-halo satellite terms separately.   Consider a spherical
halo of mass $M$ and radius $r_{180}$ that hosts a central galaxy.  It
is clear that this central  galaxy cannot contribute any signal to the
2-halo term  on scales smaller  than $r_{180}$. Hence, if  all central
galaxies  lived  in haloes  of  a  fixed  mass $M$,  then  $1+\xi^{\rm
  2h,c}_{\rm g,dm}(r)  = 0$ for  $r<r_{180}$. In reality,  though, one
needs to account for the fact  that centrals occupy haloes that span a
range  in  halo  masses,  even  if  the centrals  all  have  the  same
luminosity.  In  the case of  the satellite galaxies the  situation is
even more  complicated.  In  particular, since satellite  galaxies can
reside at  the outskirts of  dark matter haloes, the  2-halo satellite
term can  still have  non-zero power at  $r \ll r_{180}$.   Thus, halo
exclusion has  less impact  on the 2-halo  satellite term than  on the
2-halo central term.

Although  the concept  of halo  exclusion  is quite  simple, a  proper
implementation  of   it  in  the  halo  model   is  extremely  tedious
numerically. We therefore use only an approximate treatment, which has
the  advantage that it  is straightforward  to implement  numerically. 
First of all, we ignore halo  exclusion for the 2-halo satellite term. 
Since this  term is always smaller  than the 2-halo  central term, and
since  halo  exclusion  is  less  important for  satellites  than  for
centrals, this  should not have a  significant impact on  the results. 
For the 2-halo central term we proceed as follows: for each luminosity
bin, $[L_{1},L_{2}]$, we simply  set $1+\xi^{\rm 2h,c}_{\rm g,dm}(r) =
0$ for  $r < r_{180}(\bar{M})$.   Here $\bar{M}$ is the  {\it average}
halo mass of the central galaxies
\begin{equation}
\bar{M} = \int_0^{\infty} \calP_\rmc(M|L_1,L_2) \, M \, \rmd M\,,
\end{equation}
where $\calP_\rmc(M|L_1,L_2)$ is the probability that a central galaxy
with luminosity $L_1  \leq L \leq L_2$ resides in a  halo of mass $M$,
and  is given  by Eq.~(\ref{eq:prob1hc}).   The  corresponding radius,
$r_{180}(\bar{M})$,   follows   from   Eq.~(\ref{eq:mass_def}).   
\begin{figure*}
\centerline{\psfig{figure=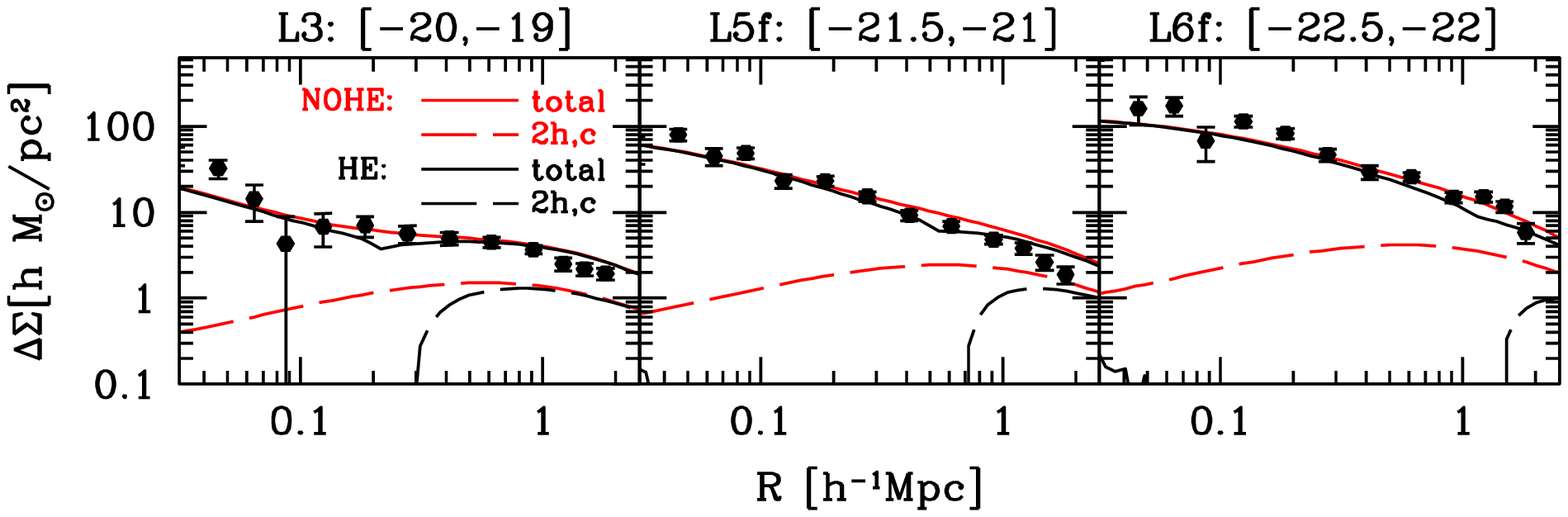,width=0.95\hdsize}}
\caption {The ESD is shown for three  luminosity bins. The black lines
  refer to  the fiducial model  (HE) and  the  red lines to  the model
  without halo  exclusion (NOHE).  The  solid lines indicate the total
  signal, whereas the long dashed lines  show the 2-halo central terms
  (note  that  the we ignore  halo  exclusion for the 2-halo satellite
  term).  Although the  2-halo  central term  is  strongly affected by
  halo exclusion, the impact on the total ESD is only mild.  Note that
  the  sharpness of  the dip  in  the black solid  lines  is (at least
  partially) an   artefact  of our   oversimplified  treatment of halo
  exclusion, as discussed in the text..}
\label{fig:AppendixPlot}
\end{figure*}

Although this treatment of  halo exclusion is  clearly oversimplified,
we emphasize  that previous attempts  to include halo exclusion in the
halo  model are  also approximations (e.g.   Magliocchetti \& Porciani
2003; Tinker \etal 2005; Yoo  \etal 2006). In  addition, as is evident
from Fig.~\ref{fig:AppendixPlot},  halo  exclusion   only has  a  mild
impact on the overall results.  The black lines, labelled HE, show the
ESDs obtained  from our  fiducial  model in   which halo exclusion  is
implemented using the method outlined  above. For comparison, the  red
lines, labelled    NOHE, show the   results in   which we  ignore halo
exclusion altogether   (i.e. in  which  the  2-halo terms  are  simply
computed  using  Eqs.\,[\ref{eq:fcP2hc}] and  [\ref{eq:fsP2hs}]).  The
dashed  lines show the corresponding  2-halo central  terms, which are
clearly suppressed on  small scales in  the HE  model.  Since brighter
central  galaxies are  hosted  by  more  massive (and therefore   more
extended)  haloes, the effect of  halo  exclusion is  apparent out  to
larger radii for brighter galaxies.  Note also  that the truncation is
fairly  sharp; this,  however,  is partially an   artefact due  to our
approximate treatment in   which   we have only  considered   the {\it
  average}  halo  mass $\bar{M}(L_1,L_2)$.    In  reality, the central
galaxies live in haloes that span a  range in halo  masses, and thus a
range in sizes. If this were to be taken  into account, the truncation
would still occur at roughly the same radius, but be less sharp.

Although  halo exclusion  clearly  has a  strong impact on  the 2-halo
central term,  the impact on the {\it  total} ESD is only modest. This
mainly owes to  the  fact that the  total  signal on small  scales  is
completely dominated by  the  1-halo terms.  Overall,  halo  exclusion
only results  in a small reduction of   the total ESD  on intermediate
scales.  Due to  the arteficial sharpness of  the break  in the 2-halo
central term, halo  exclusion introduces a sharp  feature in the total
ESD at the radius corresponding to  this break. Although the sharpness
of this feature is an artefact of our oversimplified treatment of halo
exclusion,  it  does not  influence our   overall  results.  In  fact,
including or  excluding halo exclusion has only  a small impact on the
total $\chi^2$-values  of  our  models.   For example, for   the WMAP3
cosmology,  the reduced $\chi^2$  of    our fiducial model is   $3.1$,
compared  to $4.6$  if halo  exclusion is  ignored. This difference is
much   smaller  than that  between the   WMAP1  and  WMAP3 models.  We
therefore conclude that our approximate treatment of halo exclusion is
sufficiently accurate,  and does  not impact  our conclusion  that the
WMAP3 cosmology is strongly favored over the WMAP1 cosmology.

\label{lastpage}
\end{document}

%% file: macros.tex
\newcommand{\etal}{{et al.~}}

\newcommand{\kmsmpc}{\>{\rm km}\,{\rm s}^{-1}\,{\rm Mpc}^{-1}}

\newcommand{\Mpc}{\>{\rm Mpc}}
\newcommand{\kpc}{\>{\rm kpc}}
\newcommand{\Msun}{\>{\rm M_{\odot}}}
\newcommand{\Lsun}{\>{\rm L_{\odot}}}

\newcommand{\calP}{{\cal P}}
\newcommand{\calR}{{\cal R}}

\newcommand{\beq}{\begin{equation}}
\newcommand{\eeq}{\end{equation}}

\newcommand{\rmd}{{\rm d}}
\newcommand{\rms}{{\rm s}}
\newcommand{\rmc}{{\rm c}}

\def\gtsima{$\; \buildrel > \over \sim \;$}
\def\ltsima{$\; \buildrel < \over \sim \;$}
\def\prosima{$\; \buildrel \propto \over \sim \;$}
\def\gsim{\lower.7ex\hbox{\gtsima}}
\def\lsim{\lower.7ex\hbox{\ltsima}}
\def\simgt{\lower.7ex\hbox{\gtsima}}
\def\simlt{\lower.7ex\hbox{\ltsima}}
\def\simpr{\lower.7ex\hbox{\prosima}}
\def\la{\lsim}
\def\ga{\gsim}
\def\lta{\la}
\def\gta{\ga}

\newcommand{\apj}{ApJ}
\newcommand{\apjs}{ApJS}
\newcommand{\aj}{AJ}
\newcommand{\mnras}{MNRAS}

\newdimen\hssize
\newdimen\hdsize
